\documentclass[showpacs,preprintnumbers,amsmath,amssymb]{revtex4}

\usepackage{graphicx}
\usepackage{dcolumn}
\usepackage{bm}
\usepackage{epsf}

\newcommand{\beq}{\begin{equation}}
\newcommand{\eeq}{\end{equation}}
\newcommand\beqa{\begin{eqnarray}}
\newcommand\eeqa{\end{eqnarray}}
\newcommand\bea{\begin{array}}
\newcommand\eea{\end{array}}
\newcommand\ba{\begin{array}}
\newcommand\ea{\end{array}}
\newcommand{\nn}{\nonumber}
\newcommand{\diag}[1]{{\rm diag}(#1)}
\newcommand{\neqa}{\nonumber\end{eqnarray}}
\newcommand{\la}{\label}
\newcommand{\vrho}{\varrho}
\newcommand{\vro}{\varrho}
\renewcommand{\P}{{\cal P}}
\newcommand{\p}{P}

\newcommand{\eq}[1]{eq.(\ref{#1})}
\newcommand{\eqs}[2]{eqs.(\ref{#1},\ref{#2})}
\newcommand{\Eq}[1]{Eq.(\ref{#1})}

\newcommand{\ur}[1]{(\ref{#1})}

\newcommand{\Tr}{{\rm Tr}}
\newcommand{\B}{{\cal B}}

\newcommand{\Det}{{\rm Det}}
\newcommand{\half}{\frac{1}{2}}

\renewcommand{\d}{\partial}
\renewcommand{\O}{{\cal O}}

\newcommand{\D}{r_{12}}

\renewcommand{\v}{{\rm v}}

\newcommand{\re}{\relax{\rm I\kern-.18em R}}

\newcommand{\nablaslash}{\nabla\hspace{-.65em}/\hspace{.3em}}

\def\su2{{SU(2)}}

\def\tr{{\rm tr}}

\def\s{{\rm s}}
\def\r{{\rm r}}

\begin{document}

\title{Fermionic determinant for the SU(N) caloron with nontrivial holonomy}
\begin{abstract}
In the finite-temperature Yang-Mills theory we calculate the
functional determinant for fermions in the fundamental
representation of $SU(N)$ gauge group in the background of an
instanton with non-trivial holonomy at spatial infinity. This
object, called the Kraan--van Baal -- Lee--Lu caloron, can be viewed
as composed of N Bogomolny--Prasad--Sommerfeld monopoles (or dyons).
We compute analytically two leading terms of the fermionic
determinant at large separations.

\end{abstract}
\author{Nikolay Gromov$^a$}
\email{nik_gromov@mail.ru}
\author{Sergey Slizovskiy$^a$}
\email{ssliz@list.ru}
\vskip 0.3true cm
\affiliation{$^a$ St.Petersburg INP, Gatchina, 188 300, St.Petersburg, Russia}
\maketitle
\section{Introduction}
Speaking of the finite temperature one implies that the Euclidean
space-time is compactified in the `time' direction whose inverse
circumference is the temperature $T$, with the usual periodic
boundary conditions for boson fields and anti--periodic conditions
for the fermion fields. In particular, it means that the gauge field
is periodic in time, and the theory is no longer invariant under
arbitrary gauge transformations, but only under gauge
transformations that are periodical in time. As the space topology
becomes nontrivial the number of gauge invariants increases. The new
invariant is the holonomy or the eigenvalues of the Polyakov line
that winds along the compact 'time' direction~\cite{Polyakov} \beq
L= \left.{\rm
P}\,\exp\left(\int_0^{1/T}\!dt\,A_4\right)\right|_{|\vec
x|\to\infty}. \la{Pol0}
\eeq\\
This invariant together with the topological charge and the magnetic
charge can be used for the classification of the field
configurations \cite{GPY} , its zero vacuum average is
one of the common criteria of confinement.

A generalization of the usual Belavin--Polyakov--Schwartz--Tyupkin
(BPST) instantons~\cite{BPST} for arbitrary temperatures is the
Kraan--van Baal--Lee--Lu (KvBLL) caloron with non-trivial
holonomy~\cite{KvB,KvBSUN,LL}. It is a self-dual electrically
neutral configuration with topological charge $1$ and arbitrary
holonomy. It was constructed a few years ago by Kraan and van Baal
\cite{KvB} and Lee and Lu \cite{LL} for the SU(2) gauge group and in
\cite{KvBSUN} for the general $SU(N)$ case; it has been named the
KvBLL caloron (recently the exact solutions of higher topological
charge were constructed and discussed \cite{higher}).
In the limiting case, when the KvBLL caloron is
characterized by the trivial holonomy (meaning that (\ref{Pol0})
assumes values belonging to the group center $Z(N)$ for the $SU(N)$
gauge group), it reduces to the periodic Harrington-Shepard
\cite{HS} caloron known before. It is purely $SU(2)$ configuration
and its weight was studied in detail by Gross, Pisarski and Yaffe
\cite{GPY}.

The KvBLL caloron in the theory with $SU(N)$ gauge group on the
space $R^3\times S^1$  can be interpreted as a composite of $N$
distinct fundamental monopoles (dyons)~\cite{LY}\cite{Wein} (see
fig. \ref{fig_zt} and fig. \ref{fig_zx}). It was proven in
\cite{KvBSUN} and is shown in this paper explicitly, that the exact
KvBLL gauge field reduces to a superposition of BPS dyons, when the
separation $\vrho_i$ between dyons is large (in units of inverse
temperature). When the distances $\vrho_l$ between all the dyons
become small compared to $1/T$ the KvBLL caloron reduces to the
usual BPST instanton in its core region (for explicit formulae see
\cite{KvB,DG}).

The KvBLL caloron may be relevant to the confinement-deconfinement
phase transition in the pure gauge theory \cite{DPSUSY} as well as
for the chiral restoration transition in finite-temperature QCD with
light fermions. In the latter case it is important to know the
fermionic determinant, which we calculate in this paper.

To construct the ensemble of calorons , one needs to know their
quantum weights and moduli space (zero modes). If there are massless
fermions in the theory, the ``gluonic'' quantum weight of the
caloron should be multiplied by $\left(\Det'(i
\nablaslash)\right)^{N_f}$ -- a normalized and regularized product
of fermionic non-zero modes. The fermionic zero modes would also
give a valuable contribution to interactions inside the ensemble.

Up to now, only the determinants in case of the $SU(2)$ gauge group
were found. In ref.~\cite{DGPS} the determinant for gluons and
ghosts for the $SU(2)$ Yang--Mills theory was computed. It was
extended to the SU(2) Yang-Mills theory with light fermions in
\cite{GS}. So far only a metric of the moduli space was known for
the general $SU(N)$ case \cite{Kraan} (its determinant was analyzed in
details in \cite{DG}). The fermionic zero-modes were studied in
\cite{Cherndb}. In this paper we generalize our results for the
fermionic determinant over non-zero modes to the $SU(N>2)$ gauge
group. It may be more logical to generalize the result of
\cite{DGPS} about the ghost determinant to the arbitrary $SU(N)$
first, but technically the computation of the non-perturbative
contribution of light fermions is simpler and that is why we decided
to consider it first.

\begin{figure}[t]
\centerline{
\epsfxsize=0.4\textwidth
\includegraphics[height=\epsfxsize,angle=-90]{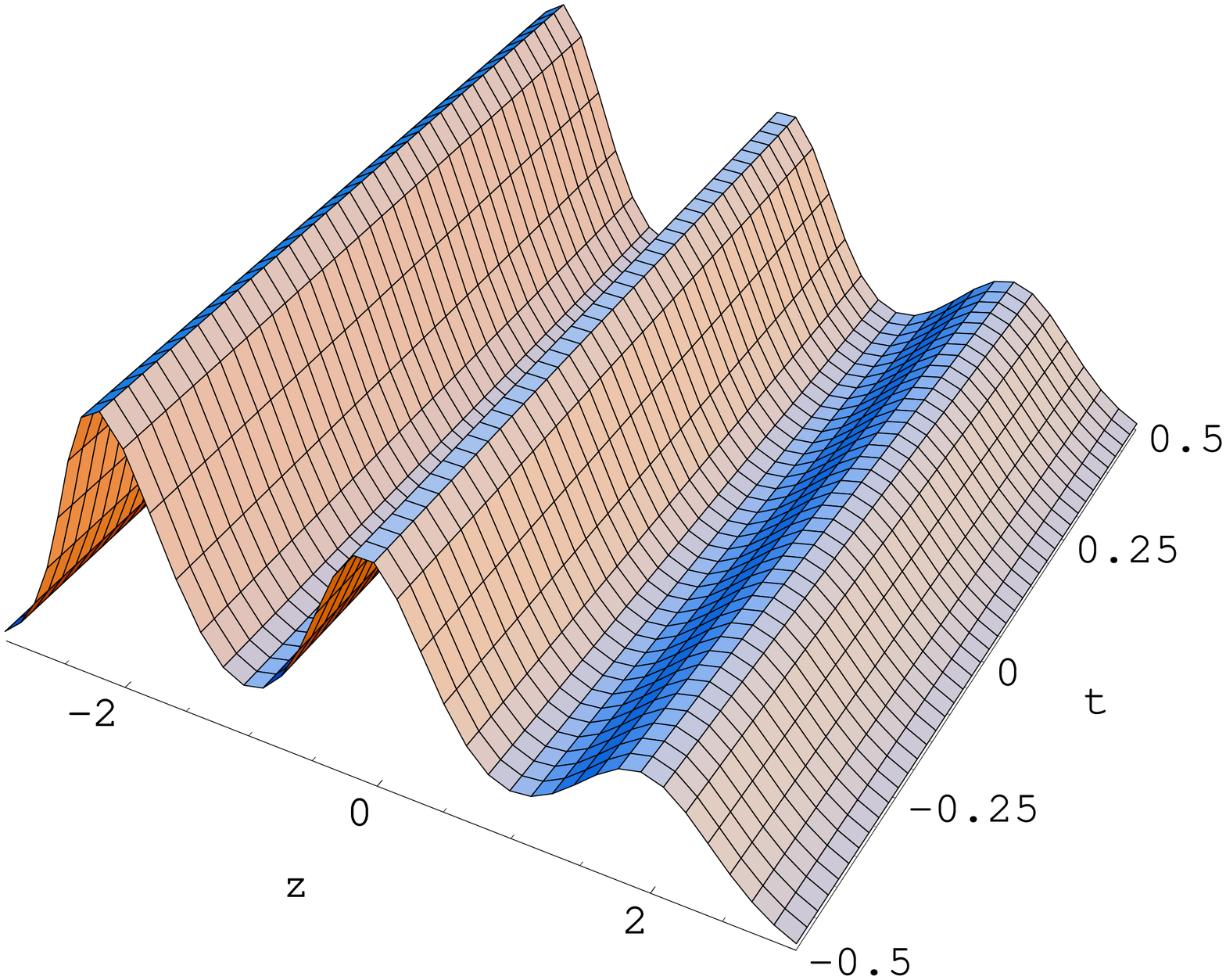}
\epsfxsize=0.4\textwidth
\includegraphics[height=\epsfxsize,angle=-90]{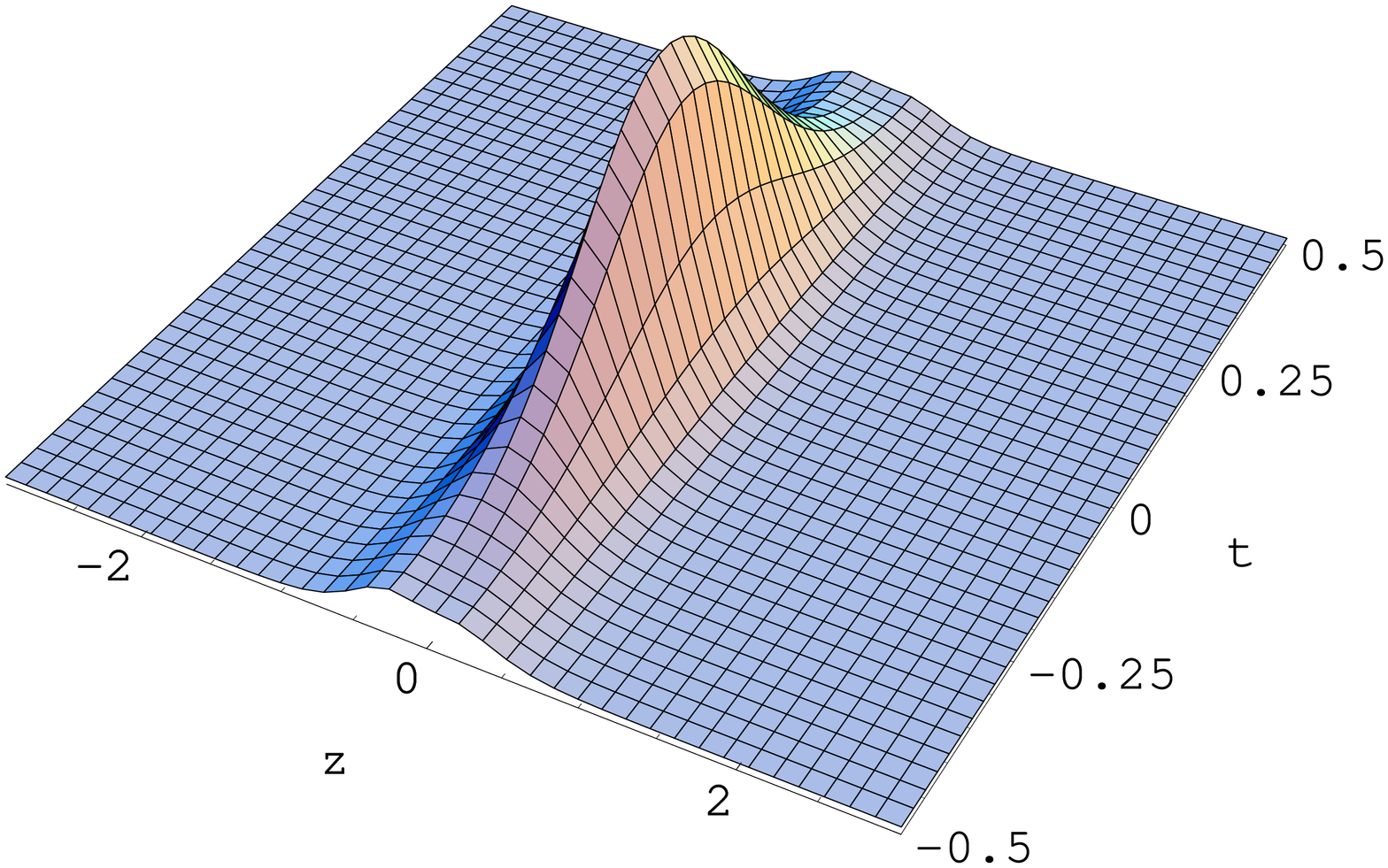}}
\caption{The action density of the $SU(3)$ KvBLL caloron as function
of $z,t$ at fixed $x=y=0$, eigenvalues of $A_4$ at spatial
infinity are $\mu_1=-0.307 T,\;\mu_2=-0.013 T,\;\mu_3=0.32 T$. It is periodic in $t$  direction.
At large dyon separation the density becomes static (left, $\vrho_{1,2}=1/T,\;\vrho_3=2/T$).
As the separation decreases the action density becomes more like a $4d$
lump (right, $\vrho_{1,2}=1/(3T),\;\vrho_3=2/(3T)$).
The axes are in units of inverse temperature $1/T$.\la{fig_zt}}
\end{figure}

Let us start a detailed exposition of our calculations. As was
already mentioned, to account for fermions we have to multiply the
partition function by $\prod\limits_{j=1}^{N_f} \Det (i \nablaslash
+i m_j)$, where $\nablaslash$ is the spin-1/2 fundamental
representation covariant derivative in the background considered,
and $N_f$ is the number of light flavors. We consider only the case
of massless fermions here $m_j=0$. The operator $i\nablaslash$ has
zero modes \cite{Cherndb} therefore a meaningful object is $\Det'(i
\nablaslash)$ --- a normalized and regularized product of non-zero
modes. In the self-dual background it is equal to $\left
(\Det(-\nabla^2) \right)^2$, where $\nabla$ is the spin-0
fundamental covariant derivative~\cite{BC}. In this work we
calculate the asymptotics of the determinant for large separations
between $N$ constituent dyons. As usual, our method of calculation
is based on calculating the variation of the determinant w.r.t. some
parameter of the solution \cite{Zar}.


 Let us sketch the structure of the paper.
To make the paper more
self-contained, in Sections \ref{notations} and \ref{ADHMcp} we
collect the notations and review the ADHMN construction of $SU(N)$
KvBLL caloron.

 A peculiar feature of fields in the fundamental representation of gauge group is that they
 feel the center elements of the group, hence there are $N$ possible different background fields,
 numbered by the integer $k=0..N-1$. They are related by a non-periodic gauge transformation (see Section \ref{difper} for detailes).
In Section \ref{reduction_sc} we discuss the $N$ possible background
fields and the boundary conditions for the fermionic fluctuations.

 In Section \ref{current_sc} we present the currents corresponding to variation of the determinant.
 Using these results we immediately write the result for the determinant
 up to an additive constant in Section \ref{determinant_sc}. To trace back the constant we shall take a special
 configuration of N far-separated constituents and will subsequently reduce
it to the $SU(2)$ configuration, where we have already calculated
the determinant in \cite{GS}. To justify this approach  we show
rigorously in Section \ref{reduction_sc} that the $SU(N)$ caloron
can be considered as a superposition of $SU(2)$ dyons  and
explicitly show how some degenerate $SU(N)$ configurations are
reduced to the $SU(N-1)$ ones. The simplicity of formulae appearing
in the main text is justified by rigorous and lengthy calculations
presented in Appendices. We also prove several statements
conjectured numerically in our previous works \cite{DGPS,GS}.

\section{Notations} \la{notations}

To help navigate and read the paper, we first introduce some notations used throughout.
Basically we use the same notations as in Ref.~\cite{KvBSUN}.
In what follows we shall measure all quantities in the temperature units and put $T=1$.
The temperature factors can be restored in the final results from dimensions.

Let the holonomy at spatial infinity have the following eigenvalues
\beq L={\rm P}\,\exp\left(\int_0^{1/T}\!dt\,A_4\right)_{|\vec
x|\to\infty} =V\,{\rm diag}\left(e^{2\pi i \mu_1},\,e^{2\pi i
\mu_2}\ldots e^{2\pi i \mu_N}\right)\,V^{-1}, \qquad
\sum_{m=1}^N\mu_m=0. \la{Pol1}\eeq
We use anti-hermitian gauge
fields $A_\mu=it^aA_\mu^a=\frac{i}{2}\lambda^aA_\mu^a$,
$[t^at^b]=if^{abc}t^c,\,\tr(t^at^b)=\half\delta^{ab}$. The
eigenvalues $\mu_m$ are uniquely defined by the condition
$\sum_{m=1}^N\mu_m=0$. If all eigenvalues are equal up to the
integer, implying $\mu_m=k/N-1,\;m\leq k$ and $\mu_m=k/N,\;m > k$
where $k=0,1,...(N-1)$, the holonomy belongs to the center of
$SU(N)$ group, and is said to be ``trivial''. By making a global
gauge rotation one can always order the holonomy eigenvalues such
that \beq \mu_1\leq \mu_2\leq \ldots \leq \mu_N\leq \mu_{N+1}\equiv
\mu_1+1, \la{mus}\eeq which we shall assume done. The eigenvalues of
$A_4$ in the adjoint representation, $A_4^{ab}=if^{abc}A_4^c$, are
$\pm(\mu_m-\mu_n)$ and $N-1$ zero ones. For the trivial holonomy all
the adjoint eigenvalues are integers. The difference between the
neighboring eigenvalues in the fundamental representation
$\nu_m\equiv\mu_{m+1}-\mu_m$ determines the spatial core size
$1/\nu_m$ of the $m^{\rm th}$ monopole whose 3-coordinates will be
denoted as $\vec y_m$, and the spatial separation between
neighboring monopoles will be denoted by
\beq \vec\vrho_m\equiv\vec
y_m-\vec y_{m-1}
=\vrho_m\,(\sin\theta_m\cos\phi_m,\,\sin\theta_m\sin\phi_m,\,\cos\theta_m),
\qquad\vrho_m\equiv |\vec\vrho_m|. \la{vrho}
\eeq
We call neighbors those dyons which correspond to the neighboring intervals in $z$ variable (see the next section),
these dyons also turn out to be neighbors in the color space.  With each
3-vector $\vec\vrho_m$ we shall associate a 2-component spinor
$\zeta^{\dagger\,\alpha}_m$ so that for any $m=1...N$: \beq
\zeta^{\dagger\,\alpha}_m\zeta^m_\beta=\frac{1}{2\pi}
\left(1_2\vrho_m-\vec\tau\cdot\vec\vrho_m\right)^\alpha_\beta\,.
\la{eq2}\eeq This condition defines $\zeta_m^\alpha$ up to $N$ phase
factors $e^{i\psi_m/2}$. These spinors are used in the construction
of the caloron field. These $\psi_m$ has the meaning of the $U(1)$
phase of the $m^{\rm th}$ dyon. For the trivial holonomy, the KvBLL
caloron reduces to the Harrington--Shepard periodic instanton at
non-zero temperatures and to the ordinary
Belavin--Polyakov--Schwartz--Tyupkin instanton at zero temperature.
Instantons are usually characterized by the scale parameter (the
``size'' of the instanton) $\rho$. It is directly related to the
dyons positions in space, actually to the perimeter of the polygon
formed by dyons, \beq \rho=\sqrt{\frac{1}{2\pi T}\sum_{m=1}^N
\vrho_m}\,,\qquad \sum_{m=1}^N \vec\vrho_m=0. \la{size}\eeq

In the next Section we shall show how the $SU(N)$ caloron gauge field depends on these
parameters and describe its ADHMN construction.
\begin{figure}[t]
\centerline{
\epsfxsize=0.4\textwidth
\includegraphics[height=\epsfxsize,angle=-90]{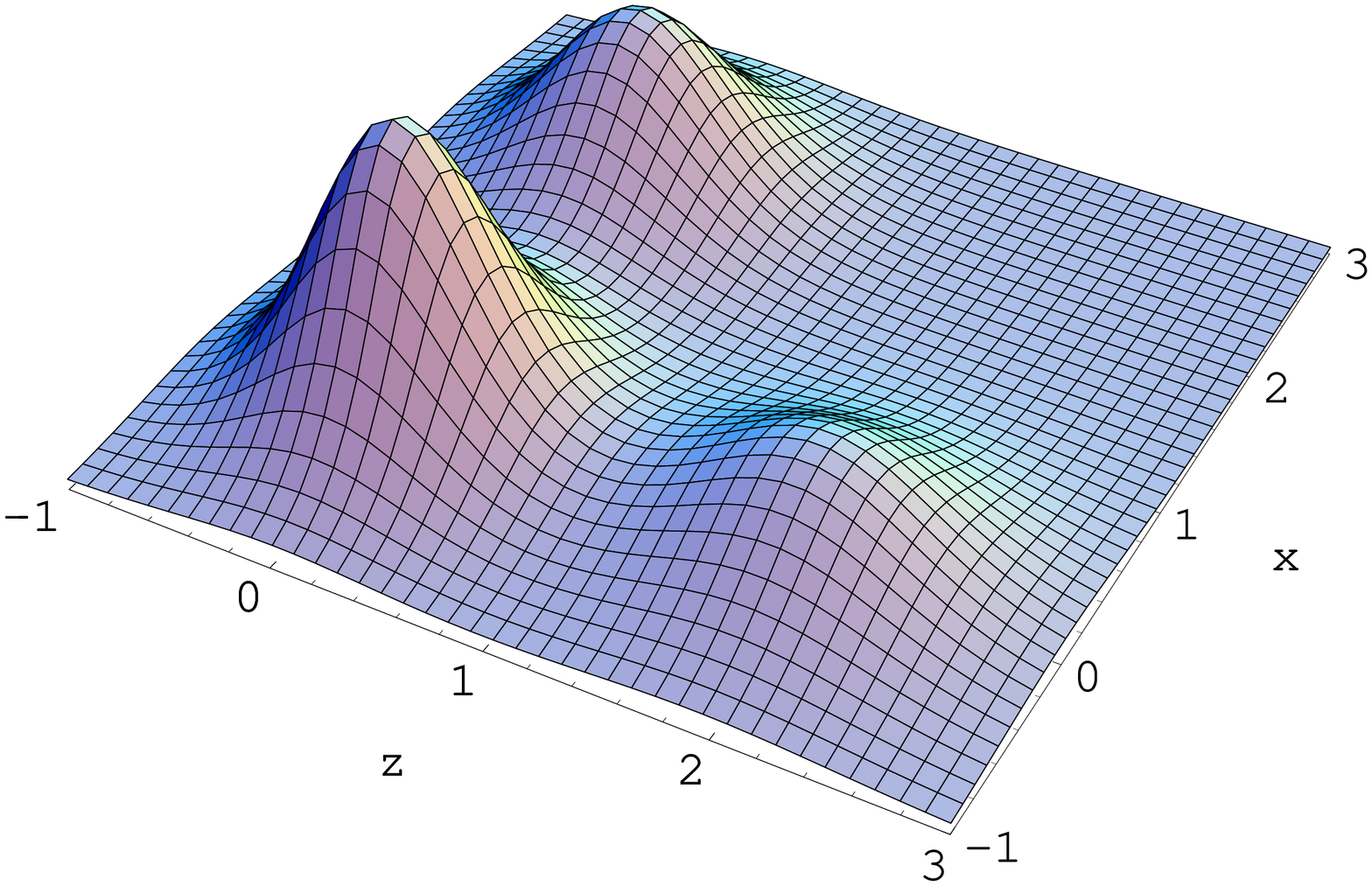}
\epsfxsize=0.4\textwidth
\includegraphics[height=\epsfxsize,angle=-90]{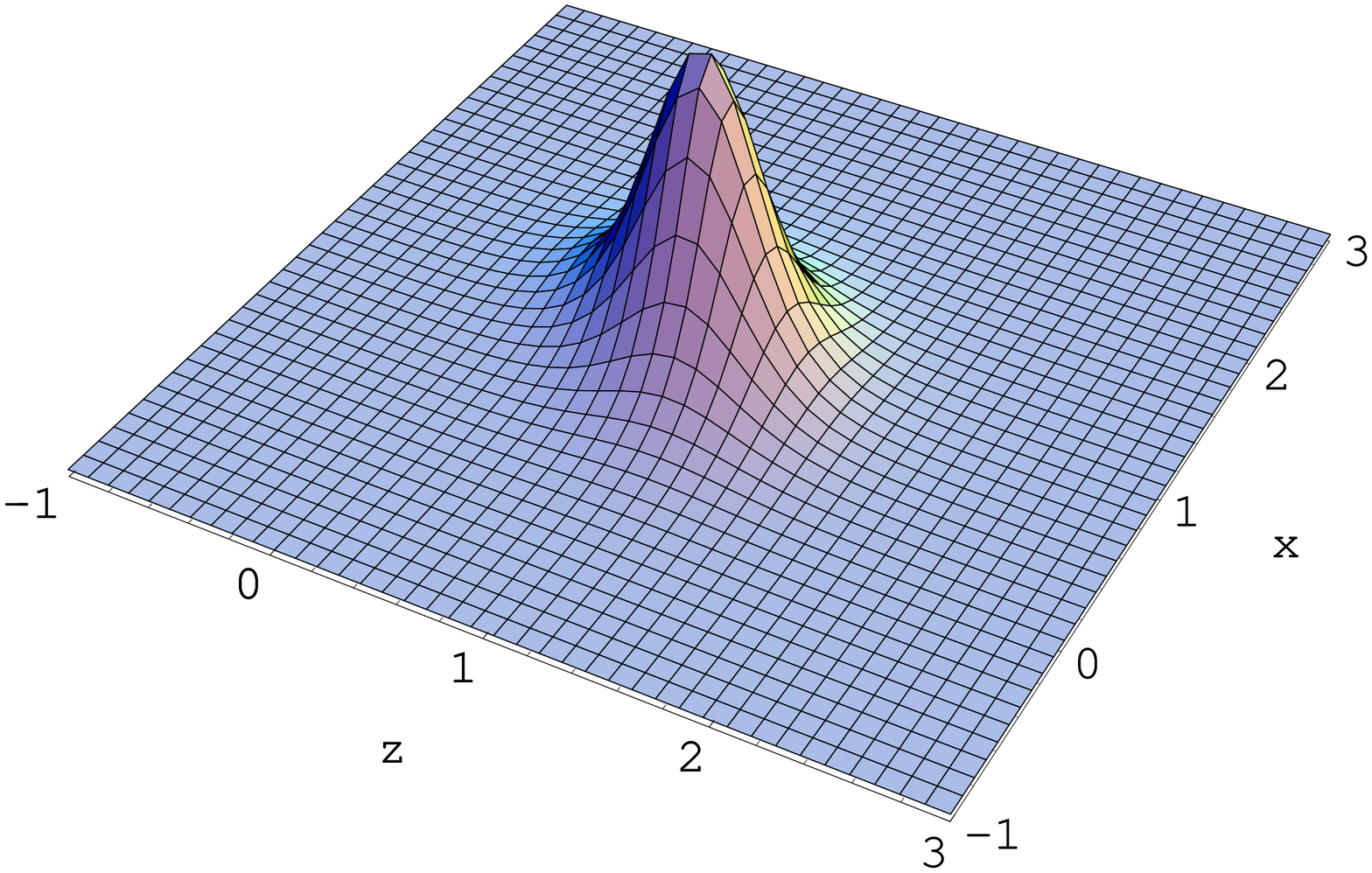}}
\caption{\la{fig_zx} The action density of the $SU(3)$ KvBLL caloron as
function of $z,x$ at fixed $t=y=0$. At large separations $\vrho_{1,2,3}$ the caloron
is a superposition of free BPS dyon solutions (left, $\vrho_1=2.8/T,\;\vrho_{2,3}=2/T$). At
small separations they merge (right, $\vrho_1=1/T,\;\vrho_{2,3}=0.54/T$). The eigenvalues of $A_0$ at spatial
infinity are the same as in Fig.~1.}
\end{figure}
\section{ADHMN construction for the SU(N) caloron }\la{ADHMcp}

Here we remind the Atiyah--Drinfeld--Hitchin--Manin--Nahm (ADHMN) construction
for the $SU(N)$ caloron~\cite{KvBSUN} and adjust it to our needs.

The basic object in the ADHMN construction~\cite{ADHM,Nahm80} is the
$(2+N)\times 2$ matrix $\Delta$ linear in the space-time variable $x$ and depending
on an additional compact variable $z$ belonging to the unit circle:
\beq
\Delta^K_\beta(z,x)=\left\{\bea{cl}
\lambda^m_\beta(z)&,\quad K=m,\quad 1\leq m\leq N,\\
(B(z)-x_\mu\sigma_\mu)^\alpha_\beta&,\quad K=N+\alpha,\quad 1\leq \alpha\leq 2,
\eea\right.
\la{Delta}\eeq
where $\alpha,\beta=1,2$ and $m=1,\dots,N$; $\sigma_\mu=(i\vec\sigma,1_2)$. As usual,
the superscripts number rows of a matrix and the subscripts number columns.
The functions $\lambda^m_\beta(z)$ forming an $N\times 2$ matrix carry information about
color orientations of the constituent dyons, encoded in the $N$ two-spinors $\zeta$:
\beq
\lambda^m_\beta(z)=\delta(z-\mu_m)\zeta^m_\beta.
\la{deflambda}\eeq
The quantities $\zeta^m_\beta$ transform as contravariant spinors of the gauge group
$SU(N)$ but as covariant spinors of the spatial $SU(2)$ group.
The $2\times 2$ matrix $B$ is a differential operator in $z$ and depends on the positions
of the dyons in the $3d$ space $\vec y_m$ and the overall position in time $\xi_4=x_4$:
\beq
B^\alpha_\beta(z)=\frac{\delta^\alpha_\beta\d_z}{2\pi i}+\frac{\hat A^\alpha_\beta(z)}{2\pi i}
\la{defB}
\eeq
with
\beq
\hat A(z)=A_\mu\sigma_\mu,\qquad \vec A(z)=2\pi i\, \vec y_m(z),\qquad A_4=2\pi i\,\xi_4,
\la{eq1}\eeq
where for $z$ inside the interval $\mu_m\leq z\leq \mu_{m+1}$, we define $\vec y(z)=\vec y_m$ to be the position
of the $m^{\rm th}$ dyon with the inverse size $\nu_m\equiv \mu_{m+1}-\mu_m$.

The gauge field of the caloron can be constructed in the following way.
One has to find $N$ quantities $v^K_n(x),\,n=1...N,$
\beq\la{v1v2}
v^K_n(x)=\left\{\bea{cl}v^{1m}_n(x)&,\quad K=m,\quad 1\leq m\leq N,\\
v^{2\alpha}_n(z,x)&,\quad K=N+\alpha,\quad 1\leq \alpha\leq 2,
\eea\right.
\eeq
which are normalized independent solutions of the differential equation
\beq\la{v1v2def}
{\lambda^\dag}^\alpha_m(z) v^{1m}_n+[B^\dag(z)-x_\mu\sigma^\dag_\mu]^\alpha_\beta v^{2\beta}_n(z,x)=0,\qquad
v^{\dag 1m}_l\,v^{1l}_n+\int_{-1/2}^{1/2}dz\,v^{\dag 2m}_\alpha\,v^{2\alpha}_n=\delta^m_n\,,
\eeq
or, in short hand notations,
\beq
\Delta^\dag v=0,\qquad v^\dag v=1_N.
\eeq
Note that only the lower component $v^2$ depends on $z$. Once $v^{1,2}$ are found,
the caloron gauge field $A_\mu$ is an anti-Hermitian $N\times N$ matrix whose
matrix elements are simply
\beq
\left(A_\mu\right)^m_n=v^{\dag 1m}_l\,\d_\mu v^{1l}_n
+\int_{-1/2}^{1/2}dz\,v^{\dag 2m}_\alpha\,\d_\mu v^{2\alpha}_n
\qquad{\rm or}\quad A_\mu=v^\dag \d_\mu v.
\la{Amu}\eeq
The gauge field is self-dual if
\beq
(\Delta^\dag\Delta)^\alpha_\beta\propto \delta^\alpha_\beta.
\la{ADHMconst}\eeq

It is important that there is a $U(1)$-internal gauge freedom.
For an arbitrary function $U(z)$, such that $|U(z)|=1$ a new operator
\beq
\Delta^K_{U,\beta}(z,x)=
\left\{
\bea{cl}
\lambda(z)^m_\beta U(z) &,\;\;\;\;\; K=m,\quad 1\leq m\leq N,\\
 U^\dag(z)(B(z)-x)^\alpha_\beta U(z) &,\;\;\;\;\; K=N+\alpha,\quad 1\leq \alpha\leq 2,
\eea\right.
\la{deltaX}\eeq
can be equally well used in the construction above.

\subsubsection{ADHM Green's function}
One can define the scalar ADHMN Green function satisfying
\beq
(\Delta^\dag\Delta)^\alpha_\beta f(z,z')=\delta^\alpha_\beta\,\delta(z-z').
\la{def_f}\eeq
From \eq{ADHMconst} one can deduce that the $N$ two-spinors $\zeta^m_\alpha$ defined
in \eq{deflambda} are associated with $\vec\vrho_m\equiv \vec y_m-\vec y_{m-1}$
according to \eq{eq2}.

\Eq{def_f} is in fact a Shr\"odinger equation on the unit circle:
\beq
\left[\left(\frac{1}{2\pi i}\d_z-x_0\right)^2+r(z)^2
+\frac{1}{2\pi}\sum_m\delta(z-\mu_m)\vro_m\right]f(z,z')=\delta(z-z')
\la{feq}\eeq
where $r(z)\equiv |\vec x-\vec y(z)|$. This equation can be solved by means of different
methods \cite{multicaloron}. We shall use the solution in the form found in \cite{DG}
\beq
f(z,z')=s_m(z) f_{mn}s^{\dag}_n(z')+2\pi s(z,z')\delta_{[z][z']}
\la{fzz}\eeq
we denoted $[z]\equiv m$ if $\mu_m\leq z <\mu_{m+1}$.
The functions appearing in \eq{fzz} are
\beqa
s_m(z)&=&e^{2\pi i x_0(z-\mu_m)}\frac{\sinh[2\pi r_m(\mu_{m+1}-z)]}{\sinh(2\pi r_m\nu_m)}\delta_{m[z]}
+e^{2\pi i x_0(z-\mu_{m})}\frac{\sinh[2\pi r_{m-1}(z-\mu_{m-1})]}
{\sinh(2\pi r_{m-1}\nu_{m-1})}\delta_{m,[z]+1},\\
s(z,z')&=&e^{2\pi i x_0(z-z')}\frac{\sinh\!\left(2\pi r_{[z]}
(\min\{z,z'\}-\mu_{[z]})\right)\sinh\!\left(2\pi r_{[z]}
(\mu_{[z]+1}-\max\{z,z'\})\right)}{r_{[z]}\sinh\!\left(2\pi r_{[z]}\nu_{[z]}\right)}.
\eeqa
In fact $s(z,z')$ is a single dyon Green's function. $N\times N$ matrix $f_{nm}=f(\mu_n,\mu_m)$ is defined by its inverse $f_{mn}={F^{-1}}_{mn}$
\beq
2\pi F_{mn}=\delta_{mn}\left[\coth(2\pi r_m\nu_m) r_m
+\coth(2\pi r_{m-1}\nu_{m-1}) r_{m-1}+\vrho_m\right]
-\frac{\delta_{m+1,n}r_m e^{-2\pi i x_0\nu_m}}{\sinh(2\pi r_m\nu_m)}
-\frac{\delta_{m,n+1} r_n  e^{2\pi i x_0 \nu_n}}{\sinh(2\pi r_n\nu_n)}\,.
\la{F}
\eeq
\Eq{fzz} is convenient since the main dependence on $z,z'$ is factorized.
Moreover a single dyon limit is manifested.

\subsubsection{Projector}
An important quantity frequently used in the ADHM calculus is the projector
\beq
P\equiv v v^\dag=1-\Delta f\Delta^\dag
\la{prj}\eeq
or, writing all indices explicitly,
\beq
P^K_L=\left\{\bea{cl}(P^{11})^m_n=v^{1m}_l\,v^{\dag 1l}_n&,\quad K=m,\quad L=n;\\
(P^{12})^m_\beta(z')=v^{1m}_l\,v^{\dag 2l}_\beta(z')&,\quad K=m,\quad L=N+\beta;\\
(P^{21})^\alpha_n(z)=v^{2\alpha}_l(z)\,v^{\dag 1l}_n&,\quad K=N+\alpha,\quad L=n;\\
(P^{22})^\alpha_\beta(z,z')=v^{2\alpha}_l(z)\,v^{\dag 2l}_\beta(z')&,\quad K=N+\alpha,\quad L=N+\beta.
\eea\right.
\la{Pdfd}\eeq
\Eq{prj} states that
\beqa
\nn (P^{11})^m_n=\delta^m_n-\zeta^m_\alpha f(\mu_m,\mu_n)\zeta^{\dag\alpha}_n,&&\quad
(P^{12})^m_\beta(z')=-\zeta^m_\alpha f(\mu_m,z')[B^\dag(z')-x^\dag]^\alpha_\beta,\\
(P^{21})^\alpha_n(z)=-[B(z)-x]^\alpha_\beta  f(z,\mu_n)\zeta^{\dag\beta}_m,&&\quad
(P^{22})^\alpha_\beta(z,z')=\delta^\alpha_\beta\delta(z-z')
-[B(z)-x]^\alpha_\gamma f(z,z')[B^\dag(z')-x^\dag]^\gamma_\beta\,.
\eeqa
It is easy to see from \eq{def_f} that both sides are projectors onto the kernel of $\Delta^\dag$:
\beq
(v v^\dag)^2=v v^\dag,\;\;\;\;\;(1-\Delta f\Delta^\dag)^2=1-\Delta f\Delta^\dag,\;\;\;\;\;{\p}v=v.
\eeq
\subsubsection{Gauge field through $f_{mn}$\la{pergauges}}

It was shown by
Kraan and van Baal~\cite{KvBSUN} that instead of \eq{Amu} one can use:
\beq
A^{mn}_\mu=\frac{1}{2}\phi_{mk}^{1/2}\zeta^k_\alpha
\bar\eta^a_{\mu\nu}(\tau^a)^\alpha_\beta\zeta^{\dag\,\beta}_l\d_\nu f_{kl}\phi_{ln}^{1/2}
+\frac{1}{2}\left(\phi^{1/2}_{mk}\d_\mu\phi_{kn}^{-1/2}-\d_\mu\phi_{mk}^{-1/2}\phi_{kn}^{1/2}\right)
\la{Amu2}\eeq
where
\beq
{\phi^{-1}}_{mn}=\delta_{mn}-\zeta^m_\alpha f_{mn}\zeta^{\dag\,\alpha}_n\,.
\la{defphi}\eeq
We see that only $f_{mn}\equiv f(\mu_m,\mu_n)$ is needed to calculate $A_\mu$.

\section{KvBLL caloron gauge field, basic features }\la{reduction_sc}
\subsection{Periodicity of the KvBLL caloron} \la{difper}
From \eq{Amu2} one can see that $A_\mu$ is \textit{not} periodical in time as it should be. More explicitly
for any integer $k$
\beq
A_\mu(x_0+k,\vec x)=g(k) A_\mu(x_0,\vec x) g^\dag(k)\la{perpr}\;
\eeq
where $g$ is a diagonal matrix $g_{mn}(k)=\delta_{mn}e^{2\pi ik\mu_n }$.
To prove (\ref{perpr}) it is enough to see from (\ref{F}) that for integer $k$
\beq
f_{mn}(k,\vec x)=f_{mn}(0,\vec x)e^{2\pi i k(\mu_m-\mu_n)}\; .
\eeq
Now we can easily make the gauge field periodic by making a time dependent
gauge transformation
\beq
A^{\rm per}_\mu=g^\dag(x_0)\d_\mu g(x_0)+g^\dag(x_0)A_\mu g(x_0)\la{Aper0}.
\eeq
However this is not the only possibility to make the field periodic in time.
Instead of $g(x_0)$ one can use $g_k(x_0)\equiv \exp[x_0\diag{2\pi i (\mu_1+k/N),\dots,2\pi i (\mu_N+k/N-1)}] $  as
$g(n)g^\dag_k(n)\in {\mathbb Z}_N$ is an element of the center of the $SU(N)$ gauge group.
Correspondingly we denote
\beq
\label{parameterk}
A^{k}_\mu=g_k^\dag(x_0)\d_\mu g_k(x_0)+g_k^\dag(x_0)A_\mu g_k(x_0)
\eeq
For different $k$, the $A^{k}_\mu$ cannot be related by a \textit{periodical}
gauge transformation. In particular the  fermionic determinant
depends explicitly on a particular choice of $k=0,\dots,N-1$. However
the expressions for $A_\mu^k$ are related as it is shown in Appendix \ref{App_gr}.
\subsection{KvBLL caloron with exponential precision}
The caloron gauge field (\ref{Amu2}) has an important feature: it is abelian
with the exponential precision, i.e. neglecting terms of the type $e^{-2\pi \mu_i r_i}$ and
$e^{-2\pi \nu_i r_i}$ one obtains \cite{DG} in the periodical gauge (\ref{Aper0})
\beqa
\la{A4ep}
A_{4\,mn}&=&2\pi i \mu_m\delta_{mn}
+\frac{i}{2}\delta_{mn}\left(\frac{1}{r_{m}}-\frac{1}{r_{m-1}}\right),\\
\vec A_{mn}&=&-\frac{i}{2}\delta_{mn}\left(\frac{1}{r_m}+\frac{1}{r_{m-1}}\right)
\sqrt{\frac{(\vrho_m-r_m+r_{m-1})(\vrho_m+r_m-r_{m-1})}
{(\vrho_m+r_m+r_{m-1})(r_m+r_{m-1}-\vrho_m)}}\;\;\vec e_{\varphi_m}\nn
\eeqa
where $\vec e_{\varphi_m}\equiv\frac{\vec r_{m-1}\times \vec r_m}{|\vec r_{m-1}\times \vec r_m|}$.

\subsection{Reduction to a single BPS dyon}
In \cite{KvBSUN} it was shown that in the domain near the $l$-th dyon where $r_l\ll r_n$ for all $n\neq l$
and the perimeter $\sum_n\vrho_n\gg 1$ is large, the action density of the KvBLL caloron reduces to
that of a single dyon (with the ${\cal O}(1/r_n)$ precision). Note that the cores of dyons may overlap and in particular
when one dyon blows up and its size $1/\nu_l$ tends to infinity all the other dyons do not
lose their shape. We will use this fact to calculate the constant in the resulting expression
for the determinant.

In Appendix \ref{reduction_to_BPS} we show explicitly how the KvBLL caloron
looks like in the vicinity of a dyon for the case of well-separated constituents
(i.e. when $e^{\nu_n r_n} \gg 1$ for all $n \neq i$ ).

\subsection{Reduction to the $SU(N-1)$ configuration} \la{reduction}
In this Section we will show that the $SU(N)$ caloron gauge field can be continuously
deformed into an $SU(N-1)$ one. This fact allows one to calculate the determinant by induction as the determinant
for the $SU(2)$ gauge group is known \cite{GS}.

Let us consider an $SU(N)$ caloron when the size of the $l$-th dyon becomes infinite (or $\nu_l=0$, meaning
$\mu_l=\mu_{l+1}$).
We shall prove that when the center of the ``disappeared'' dyon $l$ is lying on the straight line connecting
the two neighboring dyons $l-1$ and $l+1$, the resulting configuration is an $SU(N-1)$ caloron solution
having the same dyon content (except the $l$-th one) at the same positions in space.
In \cite{KvBSUN} this statement was verified for the action density. Here we show this explicitly
for the gauge field and find the gauge transformation that imbeds the $SU(N-1)$ gauge field into the upper-left $(N-1)\times (N-1)$ block
of the $SU(N)$ matrix.

It is easy to see from the definition of the
 Green's function \ur{feq} that at $\nu_l=0$ one has
$f_{ln}=f_{{l+1}n},\;f_{nl}=f_{n{l+1}}$.
 Let us denote with tilde the elements of the $SU(N-1)$ construction.
One can see from the definition \ur{feq} that
\beqa
\nn&\tilde f_{nm}=f_{nm},\;\;\;\;\;&n,m\leq l,\\
&\tilde f_{nm}=f_{n+1m+1},\;\;\;\;\;&n,m>l,\\
\nn&\tilde f_{nm}=f_{n+1m},\;\;\;\;\;&n>l,\;m\leq l,\\
\nn&\tilde f_{nm}=f_{nm+1},\;\;\;\;\;&m>l,\;n\leq l\;.
\eeqa
Since $\vec \vrho_l$ and $\vec \vrho_{l+1}$ are parallel one can write
\beqa
\nn&\tilde \zeta_{n}^\alpha=\zeta_{n}^\alpha,\;\;\;\;\;&n<l\\ \la{tzeta}
&\tilde \zeta_{n}^\alpha=\zeta_{n+1}^\alpha,\;\;\;\;\;&n>l\\
\nn&\tilde \zeta_{l}^\alpha=\sqrt{\frac{\vrho_l+\vrho_{l+1}}{\vrho_l}}\zeta_{l}^\alpha,&
\eeqa
and this is consistent with the constraint \ur{eq2}.

Let us write down explicitly the gauge transformation relating the $SU(N)$ and the $SU(N-1)$ constructions.
The crucial point is the following identity
\beq
\la{comb}\zeta^n_{\alpha} f_{nm}{\zeta^\dag}_m^\beta=(U^\dag \tilde\zeta_\alpha\tilde f\tilde\zeta^{\dag \beta} U)_{nm}
\eeq
where $U$ is a unitary matrix given by
\beqa
\nn&U_{mn}=\delta_{mn},\;\;\;\;\;&m<l\\
&U_{mn}=\delta_{mn+1},\;\;\;\;\;&m>l+1\\
\nn&U_{ln}=\delta_{ln}\sqrt{\frac{\vrho_l}{\vrho_l+\vrho_{l+1}}}-\delta_{Nn}\sqrt{\frac{\vrho_{l+1}}{\vrho_l+\vrho_{l+1}}}&\\
\nn&U_{l+1n}=\delta_{ln}\sqrt{\frac{\vrho_{l+1}}{\vrho_l+\vrho_{l+1}}}+\delta_{Nn}\sqrt{\frac{\vrho_{l}}{\vrho_l+\vrho_{l+1}}}&
\eeqa
It is assumed here that the $SU(N-1)$ construction in context of the $SU(N)$ construction is simply appended with zeroes
at the end to get the needed matrix size.
Since the gauge field \ur{Amu2} is expressed entirely through the combination (\ref{comb}), $U$ is a unitary gauge
transformation matrix that transforms $SU(N)$ configuration given by \eq{Amu2}
into $SU(N-1)$ configuration given by the same \eq{Amu2}. We see that $\zeta_l$ is consistently determined
in terms of other N-1 $\zeta$'s thus reducing by 4 the number of independent degrees of freedom.
\section{Method of computation}

In calculating the small oscillation determinant,
${\rm Det}(-\nabla^2)$, where $\nabla_\mu= \d_\mu+A_\mu$ and $A_\mu$
is the $SU(N)$ caloron field \cite{KvBSUN} in the fundamental representation,
we employ the same method as in \cite{GS,DGPS,Zar}. Instead of
computing the determinant directly, we first evaluate its derivative
with respect to a parameter $\cal P$, and then integrate the derivative
using the known determinant for the $SU(2)$ case. In case of fermions one
should consider the determinant over anti-periodical fluctuations. In Appendix \ref{App_gr} we
consider a more general problem with fluctuations periodical up
to the phase factor $e^{i\tau}$, and calculate the dependence of the determinant on $\tau$.
However for simplicity we can put $\tau=0$, i.e. consider periodical fluctuations. The dependence on
$\tau$ will be restored in the final result with the help of Appendix \ref{App_gr}. Moreover, let us note that the dependence of the
determinant on the parameter $k$ of the
gauge field (see \ur{parameterk}) is the same as the dependence on $\tau$ ($k$ can be absorbed in $\tau$ as
$\tau \to \tau+2 \pi k/N$ or vice versa).

If the background field $A_\mu$ depends on some parameter ${\cal
P}$, a general formula for the derivative of the determinant with
respect to ${\cal{P}}$ is
\beq \frac{\partial\,\log {\rm
Det}(-\nabla^2[A])}{\partial {\cal P}} = \!-\!\int
d^4x\,\Tr\left(\partial_{\cal P} A_\mu\, J_\mu\right)
\label{dvDet}
\eeq
where $J_\mu$ is the vacuum current in the
external background,  determined by the Green function: \beq
J^{ab}_\mu\!\equiv\!\left.(\delta^a_c\delta^b_d\d_x\!
-\!\delta^a_c\delta^b_d\d_y\!+\!A^{ac}\delta^b_d\!
+\!A^{db}\delta^a_c) {\cal G}^{cd}(x,y)\right|_{y= x}\qquad {\rm
or\;simply}\quad J_\mu\equiv \overrightarrow{\nabla}_\mu {\cal G}
+{\cal G} \overleftarrow{\nabla}_\mu. \label{defJ}\eeq
Here $\cal G$
is the Green's function or the propagator of spin-0, fundamental representation
particle in the given background $A_\mu$, defined by
\beq
-\nabla^2_xG(x,y)= \delta^{(4)}(x-y). \la{Gdef}\eeq
The periodic
propagator can be easily obtained from it by a standard procedure:
\begin{equation}\label{greenP}
{\cal G}(x,y)= \sum_{n= -\infty}^{+\infty} G(x_4,{\vec
x};y_4+n,{\vec y}).
\end{equation}
\Eq{dvDet} can be verified by differentiating the identity
$\log{\rm Det}(-D^2)= \Tr\log(-D^2)$.  The background field $A_\mu$
in \eq{dvDet} is taken in the fundamental representation, as is the
trace.

The Green functions in the self-dual backgrounds are generally
known~\cite{CWS,Nahm80} and are built in terms of the
Atiyah--Drinfeld--Hitchin--Manin (ADHM) construction~\cite{ADHM}
\beq
\label{green12} G(x,y)= \frac{ v^\dag(x)v(y)}
{4\pi^2(x-y)^2}\;.
\eeq

In what follows it will be convenient to split it into two parts:
\beqa \nn
{\cal G}(x,y)&=& {\cal G}^\r(x,y)+{\cal G}^\s(x,y),\\
\la{green3}
{\cal G}(x,y)^\s\equiv G(x,y),&&\qquad{\cal G}(x,y)^\r
\equiv\sum\limits_{n\neq 0}  G(x_4,{\vec x};y_4+n,{\vec
y})\la{GrGs}\;.
\eeqa
The vacuum current \ur{defJ} can be also split into
two parts, ``singular'' and ``regular'', in accordance to which
part of the periodic propagator \ur{green3} is used to calculate it:
\beq J_\mu= J^\r_\mu+J^\s_\mu. \eeq
Note that if we leave only $J^\s_\mu$ in the r.h.s. of \eq{dvDet}
then in the l.h.s. we will get a derivative of the logarithm of the determinant over
all fluctuations (not only periodical). Therefore both
\beq
\int d^4x\; \Tr(\delta A_\mu J^\s_\mu)\;\;\;\;\;{\rm and}\;\;\;\;\;\int d^4x\;\Tr(\delta A_\mu J^\r_\mu)
\eeq
are full variations of certain functional $F^\s[A],\;F^\r[A]$, such that
\beq \frac{\partial\,\log {\rm
Det}(-\nabla^2[A])}{\partial {\cal P}}=\d_{\cal P} F^\s[A]+\d_{\cal P} F^\r[A].
\eeq
By definition $F^\s[A]$ is the determinant over arbitrary fluctuations.
This defines uniquely $F^\r[A]$.
In fact $F^\r[A]$ is particularly simple and it is
calculated exactly in Appendix \ref{Appendix2}. The result is simply
\beq
F^\r[A]=\sum_n \left(P''(2\pi\mu_n)\frac{\pi\vrho_n}{4}-P'(2\pi\mu_n)\frac{\pi}{6}(y_n^2-y_{n-1}^2)+P(2\pi\mu_n)\frac{V}{2} \right),
\la{FrA}
\eeq
where $V$ is the space volume, $P(\v)=\v^2(2\pi-\v)^2/(12\pi^2)$ is the 1-loop effective potential \cite{GPY,NW}.
In the $SU(2)$ case this simple exact expression was
conjectured in \cite{GS} from numerical results. In this paper we prove it analytically for
any $SU(N)$, see Appendix \ref{Appendix2}.

As for $F^\s[A]$, we are only able to calculate this quantity
for large dyon separations. The method is the same as in \cite{DGPS,GS}.
We divide the space into the ``core'' and ``far'' domains. The first contains
well separated dyons and consists of $N$ balls of radius $R\ll 1/\nu_n$.
In the core region the r.h.s. of \eq{dvDet} is given by the simple expression computed on a
single BPS dyon with the $\O(1/\vro_n)$ precision. In the far domain the r.h.s. of \eq{dvDet} can
be computed with exponential precision. The calculations are presented in the next section.

\section{Determinant at large separations between dyons} \la{current_sc}\la{determinant_sc}
Let us consider the range of the moduli space, where the dyon cores do not overlap. To calculate
the variation of the determinant, it is convenient to divide the space into N core domains
(N balls of radius $R\gg 1/\nu_n$), and the remaining
far region. Integrating the total variation of the determinant we shall get the determinant up
to the constant that does not depend on the caloron parameters since the considered region in the moduli space is connected.
\subsection{Core domain}
In this section we calculate the r.h.s. of \eq{dvDet}
in the vicinity of the $m^{\rm th}$ dyon center. As the distances
to other dyons are large we can use simple
formulae obtained for a single BPS dyon in \cite{GS}.
We only have to make a  remark that in \cite{GS}
the calculations were made in the periodical gauge. In the present
case the gauge is not periodical. From \eq{uper} one can see
that we have to make a $U(1)$ \textit{non-periodical} gauge transformation (this results
in adding a constant proportional to a unit $2\times 2$ matrix to the BPS gauge field, see \eq{Amunear}) and
thus naively the formulae are not applicable. However in Appendix A it is shown that
only the IR-infinite terms change  under this $U(1)$ transformation
(i.e. $R$-dependent terms) and the main IR-finite part that contributes to the
caloron determinant is the same. We can conclude that the single dyon determinant depends nontrivially only
on $\nu_m=\mu_{m+1}-\mu_m$. All other changes affect only the IR-infinite terms:
\beqa
\d_{\P}\log\Det(-D^2)_{\rm near\;m^{th}\;dyon}&=&\d_\P \left(c_{dyon}\nu_m-\frac{\log(\nu_m R)}{6}\nu_m\right)+(R {\rm-dependent\; terms})
\eeqa
where $\P=\mu_n$ or $\vec y_n$.
Adding up all core contributions we obtain
\beq\la{DetCore}
\d_\P\log\Det_{{\rm core}}(-D^2)
=-\d_\P \left(\sum_n\frac{\nu_n\log(\nu_n R)}{6}\right)+(R {\rm-dependent\; terms}).
\eeq
The constant $c_{dyon}$ has disappeared here because $\sum \nu_m = 1$, and so it does not enter the  variation.
 $R$-dependent terms are exactly cancelled when we sum with the far region contribution, since the total
 result cannot depend on the choice of $R$.
\subsection{Far domain}
Now we consider the far domain, i.e. the region of space outside dyons' cores. We need to compute the vacuum current
\ur{defJ} with exponential
precision. However in fact we can obtain the result instantly using the fact that the gauge field is diagonal with
the same precision, and
for all $\mu_n\neq 0$ the Green's function \ur{greenP} falls off exponentially and thus the result can be read off from
the $SU(2)$ one.
For periodical boundary conditions we have
\beq
j_4^{mn}=\delta_{mn}\frac{i{\rm s}_m}{2}P'\left[\frac{1}{2}\left(4\pi\mu_m+\frac{1}{r_m}-\frac{1}{r_{m-1}}\right)\right],
\la{j4ep}
\eeq
where ${\rm s}_m=\frac{\mu_m}{|\mu_m|}$. All the other components are zero with exponential precision.
We have also checked this by a direct computation. It is rather involved and we do not include it in this paper.
We can immediately conclude from \eq{A4ep} for the gauge field that
\beqa\la{DetFar}
\d_\P\log\Det_{{\rm far}}(-D^2)=
\int_{far} \d_\P\frac{1}{2}\sum_n P\left[\frac{1}{2}\left(4\pi\mu_n+\frac{1}{r_n}-\frac{1}{r_{n-1}}\right)\right]\\
=\d_\P \sum_n \left(P''(2\pi\mu_n)\frac{\pi\vrho_n}{4}-P'(2\pi\mu_n)\frac{\pi}{6}(y_n^2-y_{n-1}^2)+P(2\pi\mu_n)\frac{V}{2} \right)
+(R {\rm-dependent\; terms}).\nn
\eeqa
 We have used
\beq
\int \left(\frac{1}{r_n}-\frac{1}{r_{n-1}}\right)d^3 x=\frac{2\pi}{3}(y_{n-1}^2-y_n^2),\;\;\;\;\;\int \left(\frac{1}{r_n}
-\frac{1}{r_{n-1}}\right)^2d^3 x=4\pi\vrho_{n}+(R {\rm-dependent\; terms})
\eeq
for the spherical box centered at the origin. The second equality in \ur{DetFar} is valid when the variation
does not involve changing the far region itself.
\subsection{The result}
From \eqs{DetCore}{DetFar} we can conclude that for large dyons' separations, $\vrho_m\ll 1/\nu_m+1/\nu_{m-1}$,
the $SU(N)$ caloron determinant is
\beq\la{asym}
\log\Det(-D^2)=
 \sum_n \left(P''(2\pi\mu_n)\frac{\pi\vrho_n}{4}-P'(2\pi\mu_n)\frac{\pi}{6}(y_n^2-y_{n-1}^2)+P(2\pi\mu_n)\frac{V}{2}
-\frac{\nu_n\log\nu_n}{6}\right)+c_N+\frac{1}{6}\log\mu
\eeq
where $\mu$ is the Pauli--Villars mass.
In the next section we shall show that the constant $c_N$ is the same for all $N$ and thus can be taken from the $SU(2)$ result
\cite{GS}:
 $c_N=\frac{1}{18}-\frac{\gamma_E}{6}-\frac{\pi^2}{216}+\alpha(1/2)$
where the constant $\alpha(1/2)=-\frac{17}{72}+\frac{\gamma_E}{6}+\frac{\log\pi}{6}-\frac{\zeta'(2)}{\pi^2}$ has been
introduced by 't Hooft \cite{tHooft}.

\subsection{The constant}
  We now know the {\it exact} expression \ur{FrA}
  for the regular current contribution to the variation of the determinant, and
we know the expression \ur{asym} for the determinant in the case of far dyons with
cores that do not overlap. To integrate the variation we need to know the integration constant $c_N$.
It was calculated for the $SU(2)$ case in \cite{GS}, so, to get the constant $c_N$ we will start the integration
over ${\cal P}=\nu$ from the degenerate case $\nu=0$, when the $SU(N)$ configuration is reduced to the $SU(N-1)$ KvBLL caloron.
In fact we will show that $c_N$ does not depend on $N$.

 In \cite{KvBSUN} and Section~\ref{reduction} it was shown that when two eigenvalues $\mu_l$ and $\mu_{l+1}$ of the holonomy coincide
(i.e. when the $l^{\rm th}$ dyon becomes infinitely large), and $\vec y_{l-1},\;\vec y_{l},\;\vec y_{l+1}$ belong to the same line, the $SU(N)$ configuration
 reduces to that of the $SU(N-1)$ gauge group.

The problem is that the contribution
of the singular current to the variation is not known when $\nu_l$ becomes small, because it means
that the $l^{\rm th}$ dyon overlaps the others.
We choose $\nu_1$ as a parameter $\P$ and integrate from the values of $\nu_1$ where \eq{asym} is applicable, i.e.
$\nu=\kappa/L\gg 1/L$ (we assume all $\vrho_n\sim L\gg 1$ and $\nu_{n\neq l}\sim 1$).
The problem may arise in the small region
$\nu_l \lesssim L^{-1}$ where dyons start to overlap and the integrand $\d_{\nu_1} F^\s$ is unknown.
However it is sufficient to show that
\beq
|\d_{\nu_1} F^\s| < C \log L\la{estmn}
\eeq
to prove that the contribution
from this problematic region is  small in the limit $L\rightarrow \infty$.

Again we divide all space into two parts - the core region
and the far region, but this time the core region consists of $N-1$
balls of radius $\epsilon L\gg 1/\nu_{n\neq l},\;\epsilon\ll 1$, surrounding finite size dyons.
Inside the core domain we again can use a single dyon expression for the singular contribution. It was
calculated in \cite{Zar,DGPS} and diverges logarithmically, and we can estimate it as $C\log L$.
In the far domain we can drop all terms $e^{-\nu_n r_n}$ for $n\neq l$. Let us call it the semi-exponential approximation.
As we shall show in the next paragraph,
in this domain $\d_{\nu_1} F^\s$
is a function of the form $\int d^3x\;\nu_1^3 G(r_n \nu_1,\vrho_n\nu_1)$ and thus we have to compute
\beq
\int_0^{\kappa/L} d\nu_1\int_{far} d^3 x\; \nu_1^3 G(r_n \nu_1,\vrho_n\nu_1)\la{farest}.
\eeq
To estimate this expression it is convenient to
make the following substitution:  $\vec x=L \vec x^0$, $\vec y_n=L \vec y^0_n$, $\nu_l=\nu_l^0/L$, \eq{farest} becomes
\beq
\frac{1}{L}\int_0^{\kappa} d\nu^0_1\int_{far} d^3 x^0\; {\nu^{0\;3}_1} G(r^0_n \nu^0_1,\vrho^0_n\nu^0_1)
\eeq
Since the domain of integration and the integrand do not depend on $L$,
we see that the far domain contribution tends to zero as $L\rightarrow \infty$.
Therefore only the core domain contributes, and we arrive at \eq{estmn} for large $L$.

Let us prove that $\d_{\nu_1} F^\s=-\int d^4 x \d_{\nu_1} \tr[A_\mu j^\s_\mu]$ indeed has the form $\int d^3x\;\nu_1^3 G(r_n \nu_1,\vrho_n\nu_1)$
in the semi-exponential approximation.
We can reconstruct dimensions and as the gauge field is static in this approximation,
the singular current and the gauge potential cannot depend on $T$ explicitly (as opposed to the regular current where
the temperature dependence is manifest in the definition \ur{GrGs}). It must be a spatial integral of the function of dimensionless
combinations $\nu_n\vrho_m,\;\nu_n y_m$ times $\nu_1^3$, since $F^\s$ is dimensionless. Moreover $F^s$
is independent on $\nu_{n\neq l}$ by construction. To demonstrate the latter, consider first
the gauge field. From \eq{Amu2} we see that the gauge field can be written entirely in terms
of $f_{n m}$ which by itself does not depend on $\nu_{n\neq l}$ in the semi-exponential approximation
as can be easily seen from \eq{F}.
The singular current is given by the equation (see, for example, \cite{DGPS},\cite{Zar})
\beq j^s_{\mu}= \frac{1}{12\pi^2}  v_2(z)^\dagger f(z,z')
\sigma_{\mu}(B(z')-x_\mu \sigma_\mu)^{\dagger} f(z',z'') v_2(z'') - {\rm h.c.}
\la{jsing}
\eeq
 where $v_2$ is written in \ur{u_i}, and integrations over all $z$ variables are assumed in \eq{jsing}.
 The possible $\nu_{n\neq l}$ dependence can arise from integration over $z$ the piece-wise function f
in \eq{jsing}. However $f(z,z')$ (see \eq{fzz}) is exponentially dumped as $e^{-2 \pi r_{i>1} (z-\mu_i)}$ when one
or both of its arguments are outside the interval $[\mu_{l},\mu_{l+1}]$, therefore the integrals of piece-wise functions
over these outside regions (e.g. $[\mu_{l+1},\mu_{l+2}]$) can be extended to infinity (e.g. $[\mu_{l+1},\infty)$) with
exponential accuracy. That is why no dependence on $\nu_{n\neq l}$ arises.
This completes the proof.

Thus we have shown that although \eq{asym} is valid for well separated dyons,
we can use it even when one of the dyons becomes arbitrarily large. Taking
$\mu_{l+1}=\mu_l$ and all $\vec y_{l-1},\;\vec y_{l},\;\vec y_{l+1}$ along the same line,
we see from
\beq
P''(2\pi\mu_l)\vrho_l+P''(2\pi\mu_{l+1})\vrho_{l+1}=P''(2\pi\mu_l)\tilde\vrho_l,\;\;\;\;\;
P'(2\pi\mu_l)(y_{l}^2-y^2_{l-1})+P'(2\pi\mu_{l+1})(y_{l+1}^2-y^2_{l})=P'(2\pi\mu_l)(y_{l+1}^2-y^2_{l-1})
\eeq
that \eq{asym} for $SU(N)$ reduces to that for $SU(N-1)$ with $c_{N-1}=c_{N}$.

\subsection{$\frac{\log \vrho}{\vrho}$ improvement.}
  We now calculate the first correction to the large separation
  asymptotics of the determinant (\ref{asym}). As we know from the $SU(2)$ result
  it is a $\frac{\log \vrho}{\vrho}$ correction. The correction of this special form can
  come from the far region only since the core region generates only power corrections $O(1/\vrho)$.

From \eq{DetFar} we can see that the contribution of this region is determined by the
potential energy. We take a 3d dilatation $\alpha$ (such that $\vec y_n=\alpha \vec y^0_n$) as a parameter.
We have:
\beqa
\left.\frac{\d \log
\Det(\!-\!\nabla^2)}{\d\alpha}\right|_{\rm far}
& = & \sum\limits_{n=1}^N \int\!d^3x\, \d_{\alpha} \half P\left(\half \left[4 \pi \mu_n +\frac{1}{r_n}-\frac{1}{r_{n-1}}\right]  \right). \la{smpfn}
\eeqa

The integration range for each $n$ is fixed to be the $3d$ volume with two balls ($n$-th and $n-1$-th) of radius
$R$ removed. The leading correction comes from the integral
\beqa
\int \d_{\alpha}\left(\frac{1}{r_n}-\frac{1}{r_{n-1}}\right)^4d^3x&=&
\nn\frac{32\pi\log(\vrho_n/R)}{\alpha\vrho_n}+\O(1/\D).
\eeqa
that arises when one Taylor expands $P$.
Integrating this variation over $\alpha$
we get the correction to the determinant (\ref{asym}): $ - \sum_{n=1}^N \frac{\log \vrho_n}{12 \pi \vrho_n}$.

\section{Conclusions and the final result}
  In this paper we have considered the fundamental-representation fluctuation (or fermionic) determinant over
  non-zero modes in the background field of the topological charge 1 self-dual solution at finite temperature,
  called the KvBLL caloron. This solution can be viewed as consisting of $N$ dyons.
  We have managed to calculate analytically the determinant for large dyon separations, arbitrary solution parameter $k$ (see
  (\ref{parameterk}))
  and arbitrary boundary condition for fluctuations: $$a(\vec x,1/T)=e^{-i \tau} a(\vec x,0).$$

 The result is
\beqa
\log\Det^\tau(-D^2[A^k_\mu]) &=&
 \sum_n \left(P''(2\pi\mu^{k,\tau}_n)\frac{\pi\vrho_n T}{4}-P'(2\pi\mu^{k,\tau}_n)\frac{\pi T^2}{6}(y_n^2-y_{n-1}^2)+P(2\pi\mu^{k,\tau}_n)\frac{V T^3}{2}
-\frac{\nu_n\log\nu_n}{6}-\frac{\log \vrho_n}{12 \pi \vrho_n}\right)+ \nn \\  &&+c_N+\frac{1}{6}\log\mu/T + {\cal O}(1/\vrho)
\la{final_answer}
\eeqa
where
\beq
\mu_n^{k,\tau}=\mu_n+\frac{k}{N}+\frac{\tau}{2 \pi} \;\;;\;\; c_N=-\frac{13}{72}-\frac{\pi^2}{216}+\frac{\log\pi}{6}-\frac{\zeta'(2)}{\pi^2}
\eeq
In the above expression $k = 0..N-1$ corresponds to the element of the center of the $SU(N)$ group, it
influences the result for the fundamental determinant. The anti-periodical fluctuations which are the case for fermions,
can be obtained by taking $\tau=\pi$. Therefore for the fermionic determinant $\log \Det'(i \nablaslash)$ the result is twice the
\eq{final_answer} with $\tau=\pi$.

\section*{Acknowledgements}

We thank Dmitri Diakonov and Victor Petrov for stimulating discussions.
We also thank Dmitri Diakonov for critical reading of the manuscript.  This work was partially supported by
RSGSS-1124.2003.2.

\appendix
\section{Boundary condition dependence} \la{App_gr}
In this Appendix we consider the dependence of the determinant on the boundary conditions applied for the
fluctuations around the classical configuration. For fermions they should be anti-periodical, but we consider
a more general case of
twisted boundary conditions $a(x_0=1/T) = e^{-i \tau} a(x_0=0)$.
This results in taking the twisted Green's function instead of
the periodical one \ur{greenP}:
\begin{equation}\label{greenT}
{\cal G^\tau}(A_\mu;x,y)= \sum_{n= -\infty}^{+\infty} e^{i \tau n}G(A_\mu;x_4,{\vec x},y_4+n,{\vec y}).
\end{equation}
Now let us make a $U(1)$ gauge transformation $g=e^{i\tau x_0}$. It results in adding a constant
to the fourth component of the gauge field $A_\mu\rightarrow A^\tau_\mu=A_\mu+i\tau\delta_{\mu 4}$. As
under gauge transformations the Green's function transforms as $G(x,y)\rightarrow g^\dag(x) G(x,y) g(y)$, we again
have periodical Green's function, but in a different background
\beq
{{\cal G}^\tau}(A_\mu)={{\cal G}^{\tau=0}}(A_\mu^\tau)
\eeq
Now we can consider $\tau$ as a parameter and using \eq{dvDet} we have
\beq
\d_\tau \log\Det^\tau\-\nabla^2[A_\mu] = -i\int d^4 x\; \tr\left(J_0[A_\mu^\tau]\right).
\eeq
We will calculate the r.h.s. explicitly for the general $SU(N)$ KvBLL caloron.

Only the regular current can contribute to the trace. From \eq{RegularCurrent} we have:
\beq
J^n_0=\frac{v_x^\dag v_y}{\pi^2 n^3}+\frac{v_x^\dag \B (f_x+f_y)\B^\dag v_y}{4\pi^2 n}
\eeq
where $n=y_0-x_0$. Then we will sum up $J^n_0$ with the appropriate phase factor:
\beq
4\pi^2 n^3\tr_N (J_0)=4\tr_N [v_x^\dag v_y] + n^2\tr_N[v_x^\dag \B (f_x+f_y)\B^\dag v_y].
\eeq
Let us consider the first term:
\beqa
\tr_N [v_x^\dag v_y]={v^\dag}^1_{lm}(x) v^1_{ml}(x)e^{2\pi i\mu_m n}+\int_{-1/2}^{1/2} dz\; {v^\dag}^2_{\alpha m}(x)e^{2\pi i n z}v^2_{m \alpha}(x)\\
\nn=\sum_m e^{2\pi i \mu_m n}\left(1-\frac{\varrho_m}{\pi}f_{mm}\right)+\left(2\delta(0)-\int_{-1/2}^{1/2}dz\;\tr_2(\B^\dag\Delta f\Delta^\dag\B)  e^{2\pi i n z}\right),
\eeqa
here we used \eqs{eq2}{def_f}. The notation $\delta(0)$ has the sense of the total topological charge in $\mathbb R^4$ which is infinite.
This infinity will be cancelled in a moment.

Using $[\Delta^\dag \B,e^{2\pi i n z}]=-n 1_{2}e^{2\pi i n z}$ and
\beq
\tr_2(\Delta^\dag \B\B^\dag \Delta) f(z,z')=2\delta(z-z')-\sum_m\frac{\rho_m}{\pi}\delta(z-\mu_m) f(z,z')
\eeq
we have
\beq
\tr_N [v_x^\dag v_y]=\sum_m e^{2\pi i \mu_m n}+\int_{-1/2}^{1/2}dz\;\tr_2(\B^\dag\Delta f) n\; e^{2\pi i n z}
.\eeq
The second term gives:
\beq
\tr_N[v^\dag_x \B f_x\B^\dag v_y]=\int_z\tr[\B^\dag P\B f e^{2\pi i n z}]=
\int_{-1/2}^{1/2}dz\;\left(2 f-2d_\mu f d_\mu f\right) e^{2\pi i nz}=
\int_{-1/2}^{1/2}dz\;\left[-2 f+\d_\mu (d_\mu f)\right] e^{2\pi i nz}
\eeq
($d_\mu=\frac{1}{2}\tr_2(\Delta^\dag\B\sigma_\mu)$, see the beginning of Appendix B for its properties) and we get totally
\beq
2\tr_N [v_x^\dag v_y]+n^2\tr_N[v^\dag_x \B f_x\B^\dag v_y]=
2\sum_m e^{2\pi i \mu_m n}-\int_{-1/2}^{1/2}dz\;\left[-4d_0 f_x n+2f_x n^2-\d_\mu(d_\mu f_x)n^2\right]\; e^{2\pi i n z}.
\eeq
Noting that $\int(4d_0 f_x n-2f_x n^2)e^{2\pi i n z}=\int 2n(d_0 f_x+f_x d_0)e^{2\pi i n z}=0,$ we have
\beq
\tr_N(J_0)=
\sum_m \frac{e^{2\pi i \mu_m n}}{\pi^2 n^3}+\int_{-1/2}^{1/2}dz\;\d_\mu\left[d_\mu f_x+f_x d_\mu\right]\; \frac{e^{2\pi i n z}}{4\pi^2 n}.
\eeq
And finally
\beq
\d_\tau\log\Det^\tau - \nabla^2=-i\int d^4 x\sum_{n\neq 0}
e^{i\tau n}\left(\sum_m \frac{e^{2\pi i \mu_m n}}{\pi^2 n^3}+\int_{-1/2}^{1/2}dz\;\d_\mu\left[d_\mu f_x\right]\; \frac{e^{2\pi i n z}}{2\pi^2 n}\right)
.\eeq
The first term gives $P'(2\pi\mu_m+\tau)V/2$, the second term is a full derivative and can be easily evaluated.
The exact result is
\beq
\d_\tau\log\Det^\tau - \nabla^2=P'(2\pi\mu_m+\tau)\frac{V}{2}-P''(2\pi\mu_n+\tau)\frac{\pi}{6}(y_n^2-y_{n-1}^2)+P'''(2\pi\mu_n+\tau)\frac{\pi\vrho_n}{4}
\la{dtau}
.\eeq
The main goal of this derivation is to demonstrate the technics used in the Appendix \ref{Appendix2} to prove \eq{FrA}.

Taking an appropriate limit one can deduce from \eq{dtau} how the determinant of a single dyon depends on $\tau$.
It turns out that in a dyon limit only terms proportional to $R^3,\;R^2,\;R$ in \eq{dtau} survive, where $R$ is an IR cut-off
of the space integral.

Now let us discuss the question of the relation of determinants in the different backgrounds $A^k_\mu$ (see Section \ref{difper}).
If fact $A^k_\mu$ for different $k$ can be related by the \textit{periodical} $U(N)$ gauge transformation plus
a $U(1)$ transformation $g_{U(1)}^k=e^{2\pi i x_0 k/N}$. Since the determinant does not change under periodical gauge transformations
and we know explicitly how the determinant changes under the $U(1)$ transformations we conclude that
\beq
\log\Det(-D^2[A^k_\mu])=
 \sum_n \left(P''(2\pi\mu^k_n)\frac{\pi\vrho_n}{4}-P'(2\pi\mu^k_n)\frac{\pi}{6}(y_n^2-y_{n-1}^2)+P(2\pi\mu^k_n)\frac{V}{2}
-\frac{\nu_n\log\nu_n}{6}\right)+c_N+\frac{1}{6}\log\mu\nn
\eeq
where $\mu_n^k=\mu_n+\frac{k}{N}$.
\section{Regular current, Arbitrary variation}\la{Appendix2}
In this Appendix we show that the regular part of
the determinant can be expressed as a boundary integral.
This part of the determinant results from the part of the propagator that accounts for (anti-)periodical
boundary conditions.
This fact is very important for us as it justifies
integration of our large $r_i\nu_i$ asymptotic up to the
$\nu_i=0$, and thus reduce $SU(N)$ caloron to $SU(N-1)$
to obtain a constant.

Our considerations are general and can be
generalized to arbitrary topological charges
since we use only the general properties of the ADHM
construction for periodical configurations.

Let us introduce the following notations
\beq
e\equiv e^{-2\pi i n z},\;\;\;\;\;d_\mu=\frac{1}{2}\tr_2(\Delta^\dag\B\sigma_\mu).
\eeq
 Using the ADHM constraint \eq{ADHMconst}, one can show that $d_\mu$ is a Hermitian operator.
In case of one caloron we simply have $d_\mu=r_\mu(z)-\delta_{\mu 0}\frac{\d_z}{2\pi i}$.
It can be shown that
\beq
\d_\mu f =-2 f d_\mu f,\;\;\;\;\;\d_\mu d_\nu=\delta_{\mu\nu}.
\eeq
It is straightforward to show that
\beqa
[d_\mu,e]=\delta_{\mu 0} e n,\;\;\;\;\;[f,e]=f[e,d_\mu d_\mu]f,
\eeqa
to derive the last equality we have used that $[f,{e}]=f[{e},f^{-1}]f$.
Denoting $v=v(x), v_y=v(y)$ where $x_i=y_i, x_0=n+y_0$, we have
\beqa
vv^\dag v_y v^\dag=P v_y v^\dag=v_y v^\dag-\Delta f n\B^\dag v_y v^\dag=v_y v^\dag-\Delta f n e\B^\dag P
\eeqa
where $P=(1-\Delta f\Delta^\dag)=vv^\dag$. In the last equality we have used the periodicity property of $v$:
$\B v_y=e\B v$. Now consider an expression for the regular current:
\beq
J^{\rm r}_\mu=D^x_\mu \frac{v^\dag v_y}{4\pi^2 (x-y)^2}+\frac{v^\dag v_y}{4\pi^2 (x-y)^2}\overleftarrow{D}^y_\mu
=-\frac{v^\dag v_y}{\pi^2 n^3}-\frac{v^\dag\B (f e+ef)\B v}{4\pi^2 n}\equiv J^1_\mu+J^2_\mu,
\la{RegularCurrent}
\eeq
to obtain this representation we have used the periodicity property of the ADHM Green's function $f_y=e f_x e^\dag$.
The variation of the gauge field can be expressed in the following way \cite{KvB,DG}
\beq
\delta A_\mu =D_\mu\left( v^\dag \delta v\right)
+\left(v^\dag\delta{\Delta}f\d_\mu\Delta^\dag v-v^\dag\d_\mu\Delta f\delta{\Delta}^\dag v\right)
\equiv \delta A^1_\mu+\delta A^2_\mu.
\eeq
 As far as $v^\dag \delta v$ is periodic in time, we can drop the first term due
to the current conservation (otherwise the boundary term appears).
Using the above identities and notations we have
\beqa
\la{A2J1}
-\pi^2n^3\delta A^2_\mu J^1_\mu
&=&(\delta\Delta f \B^\dag-\B f\delta\Delta^\dag)
\left(v_y v^\dag-\Delta f e n\B^\dag P\right)\\ \nn
&=&\delta\Delta f e\B^\dag P-f\delta\Delta^\dag P_y \B e
-\delta\Delta f \B^\dag \Delta f e n\B^\dag P+\B f\delta\Delta^\dag\Delta f e n\B^\dag P\\ \nn
&=&\left(\B^\dag\delta\Delta f e-f\delta\Delta^\dag\B e
-\B^\dag\delta\Delta f \B^\dag \Delta f e n+f\delta\Delta^\dag\Delta f e n\right)\\ \nn
&+&\left(-\Delta^\dag\delta\Delta f e\B^\dag \Delta f + f\delta\Delta_y^\dag \Delta_y f_y\Delta_y^\dag \B e
+ \Delta^\dag\delta\Delta f \B^\dag \Delta f e n\B^\dag \Delta f -\Delta^\dag\B f\delta\Delta^\dag\Delta f e n\B^\dag  \Delta f\right)
\eeqa
(we assume all indexes to be contracted, in particular, we do not always write the trace over spinor space).
It is convenient to denote
\beq
M_\mu=\frac{1}{2}\tr(\delta\Delta^\dag\Delta\sigma_\mu^\dag),\;\;\;\;\;\d_\nu M_\mu=\frac{1}{2}\tr(\delta\Delta^\dag\B\sigma_\nu\sigma_\mu^\dag)
.\eeq
This operator can always be made hermitian by the internal $U(1)$ gauge transformation.
We shall assume $M_\mu$ to be Hermitian in this Section. In the next Section we shall
write $M_\mu$ explicitly for certain variations. With the help of the new notations we
can proceed with eq.(\ref{A2J1})
\beqa
\la{A2J12}
&&2 f M_0 f en-2 \d_0 M_\mu f d_\mu f en+2 M_\mu f [d_\mu f, e]
+n M_\mu f d_\nu f e d_\alpha f\tr(\sigma_\mu^\dag\sigma_\nu\sigma_\alpha)
-n M_\mu f e d_\nu f d_\alpha f\tr(\sigma_\mu\sigma_\nu\sigma_\alpha^\dag)
\\ \nn
&=&4 f M_0 f en-2 M_0 f\{ d_\mu f, en\} d_\mu f+\d_\mu(\d_0 M_\mu f)en
+2 M_\mu f [en,d_\mu f d_0] f-2\epsilon_{\mu\nu\alpha}M_\mu f[d_\nu f,en]d_\alpha f+2 M_\mu f[d_0 f,en]d_\mu f
\\ \nn
&=&\d_\mu(f M_0 f d_\mu)en+\frac{1}{2}\d_\mu(\d_\mu M_0 f)en+\d_\mu(\d_0 M_\mu f)en+A
.\eeqa
We have used $f\delta\Delta_y^\dag \Delta_y f_y\Delta_y^\dag \B e=f e\delta\Delta^\dag \Delta f\Delta^\dag \B$ and denoted
\beqa \la{Adef}
 A\equiv\overbrace{2 f d_\mu f M_\mu f en^2-2 f M_\mu f d_\mu f en^2}^{\Lambda^{f^3}}
+2(\stackrel{\Lambda^{f^4}}{\overbrace{\delta_{\nu 0}\delta_{\mu\alpha}-\delta_{\alpha 0}\delta_{\nu\mu}}}-\stackrel{A_{11}}{\overbrace{\epsilon_{\mu\nu\alpha})
f d_\alpha f M_\mu f d_\nu f [e,d^2]n}}+\\ \underbrace{\d_\mu M_0 f d_\mu f [e,d^2] f n}_{\Lambda^{f^4}}
\nn+\underbrace{2 f M_0 f d_0 f en^2}_{\Lambda^{f^3}}.
\eeqa
Note that terms which are not full derivatives in eq.(\ref{A2J12}) are of order $n^2$. As we
shall see they cancel exactly with contributions coming from $J^2$:
\beq
-4\pi^2 n\; \tr_N(\delta A_\mu^2 J_\mu^2)=B_1+B_2
,\eeq
where the first contribution is
\beqa
\nn B_1&=&(\delta\Delta f \sigma_\beta^\dag\B^\dag-\B\sigma_\beta f\delta\Delta^\dag)\Delta f\Delta^\dag
\B (\sigma_\beta f e+e f \sigma^\dag_\beta)\B^\dag \Delta f\Delta^\dag\\
\la{B1def}&=&\stackrel{B_{11}}{\overbrace{4(M_\mu f d_\nu f d_\nu [f,e] d_\mu f+M_\mu f d_\nu f [d_\nu f,e] d_\mu f)+
4(M_\mu f d_\mu [f d_\nu, e] f d_\nu f+M_\mu f d_\mu [f,e] d_\nu f d_\nu f)}}\\ \nn
&+&\underbrace{4 M_\mu f d_\nu f [e,d_\mu] f d_\nu f}_{B_{11}}
-\underbrace{4 M_\mu f d_\nu f \{d_\alpha,e\} f d_\beta f\epsilon_{\mu\nu\alpha\beta}}_{B_{12}}\;,
\eeqa
the second contribution is
\beqa
\nn B_2&=&-(\delta\Delta f \sigma_\beta^\dag\B^\dag-\B\sigma_\beta f\delta\Delta^\dag)\Delta f\Delta^\dag
\B (\sigma_\beta f e+e f \sigma^\dag_\beta)\B^\dag
-(\delta\Delta f \sigma_\beta^\dag\B^\dag-\B\sigma_\beta f\delta\Delta^\dag)
\B (\sigma_\beta f e+e f \sigma^\dag_\beta)\B^\dag \Delta f\Delta^\dag\\ \la{B2def}
&=&\stackrel{\Lambda^{f^4}}{\overbrace{4 \d_0 M_\mu f \{ d_0 f d_\mu,e\} f-4 \d_0 M_\mu f \{ d_\mu f d_0,e\} f}}
-\stackrel{0}{\overbrace{4\epsilon_{\mu\nu\alpha} \d_0 M_\mu f \{ d_\nu f d_\alpha, e\} f}}\la{B2}\\
&-&\underbrace{8 f e d_\mu f M_\mu f+4 f d_\mu f M_\mu f e+8 f M_\mu f d_\mu e f-4 f M_\mu f d_\mu f e}_{\Lambda^{f^3}}.
 \nn
\eeqa
We denote $\epsilon_{\alpha\beta\gamma}\equiv\epsilon_{0\alpha\beta\gamma}$. We shall use the additional assumption that
$\d_0 M_0=0$ or, equivalently, $\delta d_0=0$.
For certain variations this condition is consistent with the requirement of hermisity of $M_\mu$ as we shall
see in the next subsection. Important consequences of this
assumption are $\d_\mu M_\nu=-\d_\nu M_\mu, \d_\mu M_\mu=0$ and $\d_\nu M_\mu \epsilon_{\mu\nu\alpha\beta}=2 \d_\alpha M_\beta$.
Let us demonstrate that the third term in eq.(\ref{B2}) is zero
when integrated over $x_0$. Let us use the following
properties of the ADHM construction of the caloron:
\beq
f(z,z',x_0)=f(z',z,-x_0)\;\;\;\;\;{\rm or}\;\;\;\;\;f(x_0)=f^T(-x_0).
\eeq
Similarly
\beq
d^T_i=d_i,\;\;\;\;\;d_0^T(x_0)=-d_0(-x_0),\;\;\;\;\;(\d_0 M_\mu)^T=\d_0 M_\mu,
\eeq
\beq
\eta_{\mu\nu}\epsilon_{\mu\nu\alpha\beta}=-2\eta_{\alpha\beta},\;\;\;\;\;\d_\nu M_\mu \epsilon_{\mu\nu\alpha\beta}=2 \d_\alpha M_\beta
\eeq
then from the simple fact that $\tr(N)=\tr(N^T)$ we have (in our case "$\tr$" means integration over $z$)
\beq
\int_{-1/2}^{1/2}\left(\epsilon_{\mu\nu\alpha} \d_0 M_\mu f \{ d_\nu f d_\alpha, e\} f \right)dx_0=
\int_{-1/2}^{1/2}\left(\epsilon_{\mu\nu\alpha} \d_0 M_\mu f d_\nu f d_\alpha e f
+\epsilon_{\mu\nu\alpha} (\d_0 M_\mu)^T f^T d^T_\alpha f^T d^T_\nu e f^T \right)dx_0=0
.\eeq
In what follows we frequently use the trick like this. We do not write integral over $x_0$ explicitly
but always assume it.

Consider the term marked by $A_{11}$ in (\ref{Adef}):
\beqa
\frac{4A_{11}}{n} &=&8 M_\mu f d_\nu[e,f]d_\alpha f \epsilon_{\mu\nu\alpha}
=4\d_\nu (M_\mu f d_\alpha e f)\epsilon_{\mu\nu\alpha}-4\d_\nu M_\mu f d_\alpha e f\epsilon_{\mu\nu\alpha}\\ \nn
&=&4\d_\nu (M_\mu f d_\alpha e f)\epsilon_{\mu\nu\alpha}-8\d_0 M_\mu f d_\mu e f=
4\d_\nu (M_\mu f d_\alpha e f)\epsilon_{\mu\nu\alpha}+4\d_\mu(\d_0 M_\mu f e)-8\d_0 M_\mu f d_\mu [e, f]
.\eeqa

Consider the term marked by $B_{12}$ in eq.(\ref{B1def}):
\beqa
B_{12}&=&-4 M_\mu f d_\nu f\{d_\alpha,e\}f\d_\beta f\epsilon_{\mu\nu\alpha\beta}
=\d_\nu(M_\mu f\{ d_\alpha,e\} f d_\beta f)\epsilon_{\mu\nu\alpha\beta}
+\d_\beta(M_\mu f d_\nu f \{ d_\alpha,e\} f)\epsilon_{\mu\nu\alpha\beta}
\\ &+&\nn4\d_\mu M_0 f d_0 f d_\mu e f-4\d_\mu M_0 f d_\mu e f d_0 f
+2 \d_\mu M_0 f \{d_0,e\}f d_\mu f
-2 \d_\mu M_0 f d_\mu f \{d_0,e\} f
.\eeqa
Combining them we have
\beqa
\la{A11B12}\frac{4A_{11}}{n^2} \!\!+\!B_{12}\!&\simeq&\!
-\!\frac{8}{n}\d_0 M_\mu f d_\mu [e, f]\!+\!4\d_\mu M_0 f d_0 f d_\mu e f\!-\!4\d_\mu M_0 f d_\mu e f d_0 f
\\ \nn
\!&+&\!2 \d_\mu M_0 f \{d_0,e\}f d_\mu f
\!-\!2 \d_\mu M_0 f d_\mu f \{d_0,e\} f\!=\underbrace{4 \d_\mu M_0 f [d_0 f,e]d_\mu f+4 \d_\mu M_0 f d_\mu [f,e]d_0 f}_{\Lambda^{f^4}}
.\eeqa
The sign "$\simeq$" means that the equality is valid up to a full derivative.
We shall collect the full derivatives at the  end of this calculation.

It is straightforward to check that all terms in $B_1$ marked by $B_{11}$ can be expressed
in the following form
\beqa
\la{B11def}B_{11}&=&
-2\d_\nu(M_\mu f d_\nu f e d_\mu f)+2\d_\nu(M_\mu f d_\mu e f d_\nu f)+
2\d_\nu(M_\mu f d_\nu e f d_\mu f)-2\d_\nu(M_\mu f d_\mu  fe d_\nu f)\\ \nn
&-&2\d_\mu(M_\mu f d_\nu e f d_\nu f)+2\d_\mu(M_\mu f d_\nu fe  d_\nu f)+
2 \d_\nu M_\mu  f d_\nu [f,e] d_\mu f+2 \d_\nu M_\mu f d_\mu [f,e] d_\nu f\\ \nn
&+&8 f M_\mu f d_\mu[f,e]+8 f M_\mu f [f,e] d_\mu
\simeq
\underbrace{8 f M_\mu f d_\mu[f,e]+8 f M_\mu f [f,e] d_\mu}_{\Lambda^{f^3}}
.\eeqa

Combining all terms in eqs.(\ref{Adef},\ref{B2def},\ref{A11B12}) marked by $\Lambda^{f^4}$ we have
\beqa
\la{deflf4}\Lambda^{f^4}&=&-\frac{8}{n}(\delta_{\nu 0}\delta_{\mu\alpha}-\delta_{\alpha 0}\delta_{\mu\nu})M_\mu f d_\nu [e,f]d_\alpha f
-\frac{4}{n} \d_\mu M_0 f d_\mu[e,f]+4 \d_0 M_\mu f \{ d_0 f d_\mu,e\} f-4 \d_0 M_\mu f \{ d_\mu f d_0,e\} f\\ \nn
&-&4 \d_0 M_\mu f[d_0 f,e]d_\mu f-4 \d_0 M_\mu f d_\mu [f,e]d_0 f=
8M_\mu f d_\mu f\{d_0,e\}fd_0 f-8M_\mu f d_0 f \{d_0,e\}fd_\mu f+8 \d_0 M_\mu f e d_0 f d_\mu f\\ \nn
&=&8\d_0(M_\mu f e d_0 f d_\mu f)-8 \d_\mu(M_\mu f e d_0 f d_0 f)+8 M_\mu f d_\mu f e f d_0 f n-8 M_\mu f d_0 f e f d_\mu f n
-8 f d_i f M_i f e\\ \nn
&=&-8 \d_\mu(M_\mu f e d_0 f d_0 f)-2\d_\mu(M_\mu f e f d_0 f)n+2\d_\mu(M_\mu f d_0 f e f)n
-8 f d_i f M_i f e\simeq-\underbrace{8 f d_i f M_i f e}_{\Lambda^{f^3}}
.\eeqa
Combining all terms in eqs.(\ref{Adef},\ref{B2def},\ref{A11B12},\ref{B11def},\ref{deflf4}) marked by $\Lambda^{f^3}$ we have
\beqa
\Lambda^{f^3}&=&(8 f d_\mu f M_\mu f e-8 f M_\mu f d_\mu f e+8 f M_0 f d_0 f e)-8 f d_i f M_i f e
-8 f e d_\mu f M_\mu f+4 f d_\mu f M_\mu f e\\ \nn
&+&8 f M_\mu f d_\mu e f- 4 f M_\mu f d_\mu f e+
8 f M_\mu f d_\mu f e-8 f M_\mu f d_\mu e f+8 f e d_\mu f M_\mu f-8 f d_\mu f M_\mu f e\\ \nn
&=&4 f d_\mu f M_\mu f e-4 f M_\mu f d_\mu f e+8 f M_0 f d_0 f e-8 f d_i f M_i f e=2\d_\mu(f M_\mu f)e
.\eeqa
Finally we collect all the full derivative terms and the result is the following expression,
which is a full derivative
\beqa
\nn-4 \pi^2 n\;\delta A^2_\mu J^{\rm r}_\mu&=&
\frac{1}{n} \left[4\d_\mu(\d_0 M_\mu f)e +4\d_\mu(f M_0 f d_\mu)e +2\d_\mu(\d_\mu M_0 f)e+
4\d_\nu(M_\mu f d_\alpha e f)\epsilon_{\mu\nu\alpha}+4\d_\mu(\d_0 M_\mu f e)\right]
\\ \la{maineq}
&+&\d_\nu(M_\mu f\{d_\alpha,e\}f d_\beta f)
\epsilon_{\mu\nu\alpha\beta}+\d_\beta(M_\mu f d_\nu f \{ d_\alpha,e\} f)\epsilon_{\mu\nu\alpha\beta}
\\ \nn
&-&2\d_\nu(M_\mu f d_\nu f e d_\mu f)+2\d_\nu(M_\mu f d_\mu e f d_\nu f)+
2\d_\nu(M_\mu f d_\nu e f d_\mu f)-2\d_\nu(M_\mu f d_\mu  fe d_\nu f)
\\ \nn
&-&2\d_\mu(M_\mu f d_\nu e f d_\nu f)+2\d_\mu(M_\mu f d_\nu fe  d_\nu f)
-8 \d_\mu(M_\mu f e d_0 f d_0 f)-2\d_\mu(M_\mu f e f d_0 f)n
\\ \nn
&+&2\d_\mu(M_\mu f d_0 f e f)n+2\d_\mu(fM_\mu f)e.
\eeqa
The fact that the result is a full derivative means that the exponential precision turns out to be exact
for this part of the determinant. This phenomenon was first discovered numerically for the $SU(2)$
gauge group. For the case of the trivial holonomy the similar fact was noticed by \cite{GPY}.

Now let us choose the parameter of variation.
We define $\vec y_i=\alpha \vec y^0_i$ and vary with respect to $\alpha$. It is easy to see that
\beqa
\Delta^\dag\delta\Delta=\frac{1}{4\pi}(\vrho_n-\vec\vro_n\vec\tau_n)\delta(z-\mu_n)+\left(\frac{\d_z}{2\pi i}-r^\dag(z)\right)y(z),\\
\delta\Delta^\dag\Delta=\frac{1}{4\pi}(\vrho_n-\vec\vro_n\vec\tau_n)\delta(z-\mu_n)+y^\dag(z)\left(\frac{\d_z}{2\pi i}-r(z)\right).
\eeqa
One can see that $M$ is hermitian
and that $\d_0 M_0=0$
\beqa
M_0&=&\frac{\vrho_n}{4\pi}\delta(z-\mu_n)-\vec r(z)\cdot \vec y(z),\\
\vec M&=&-\frac{\vec \vrho_n}{4\pi i}\delta(z-\mu_n)+[\vec r(z)\times\vec y(z)]-\vec y(z)\left(\frac{\d_z}{2\pi i}-x_0\right)\;.
\eeqa
We can substitute this $M_\mu$ to the main \eq{maineq}. As the expression is a full derivative, it is enough to evaluate
it with the exponential precision. It turns out that only the terms in the first line contribute to the boundary integral and the r.h.s.
of \eq{maineq} can be easily evaluated to give
\beq
\d_\alpha F^\r[A]=\sum_n \left(P''(2\pi\mu_n)\frac{\pi\vrho_n}{4}-P'(2\pi\mu_n)\frac{\pi}{3}(y_n^2-y_{n-1}^2) \right)
\la{FrA1}.
\eeq
As $\alpha$ goes to zero the KvBLL caloron reduces to the usual zero temperature instanton plus a constant
field. The instanton is point-like and does not give a contribution to $F^\r$, hence for zero $\alpha$ we have
\beq
F^\r(\alpha=0)=\sum_n P(2\pi\mu_n) \frac{V}{2}.
\eeq
Integrating \ur{FrA1} over $\alpha$ we come to \eq{FrA}. This result is in a perfect agreement with our $SU(2)$ results \cite{GS} and with
derivation of asymptotic of large separations \eq{asym}.

\section{Reduction to a single dyon }\la{reduction_to_BPS}
In this Appendix we shall show that
the $SU(N)$ KvBLL caloron gauge field reduces to the field of BPS dyon situated at the point
$r_l=0$ with a topological charge $\nu_l$. It happens in
the domain of the caloron moduli space $\vrho_l,\vro_{l+1}\gg 1/\nu_l$
and in the domain of space-time $r_l\lesssim 1/\nu_l$
i.e. near the point where the dyon is situated.
Without loss of generality we can consider $l=1$.

We denote $\varrho_{1,2}=\rho_{1,2} R$, where $R$ is now a large
expansion parameter. For simplicity we also assume all $|\vec y_i-\vec y_j|$ to be large,
we discuss the other possibilities
at the end of this Appendix.
For our goal it is enough to show how the ADHMN construction
of a caloron becomes that of a dyon. We find the  leading  in $R$ term of $v$ from \eq{v1v2}.
We will see that $v^1=\O(1/R)$ and $v^2$ coincides with corresponding dyon's $v_{\rm dyon}$.

First we consider a Green's function $f(\mu_n,\mu_m) \equiv f_{nm}$.
As it follows directly from \eq{F} the matrix $f_{nm}$ is diagonal with exponential accuracy except for the
upper-left $2 \times 2$ block,  which is diagonal only up to the $1/R$ terms. With the $1/R^2$ accuracy,
the  non-zero elements of $f_{mn}$ are
\beqa
\nn f_{11}&=& \frac{\pi}{R \rho_1}-\frac{ \pi
        \left( \vec r \cdot \vec \rho_1 +
          r \rho_1 \coth (2 \pi r \nu) \right)}{2 R^2 {\rho_{1}}} ,\\
 f_{22}&=& \frac{\pi}{R \rho_2}-\frac{ \pi
        \left( \vec r \cdot \vec \rho_2 +
          r \rho_2 \coth (2 \pi r \nu) \right)}{2 R^2 {\rho_{2}}} ,\\
\nn f_{12}&=&f_{21}^\dag= \frac{\pi r e^{-2 \pi i
         x_0 \nu}}{2
       \sinh(2 \pi r
        \nu_1) R^2 \rho_{1}
      \rho_{2}} ,\\
\nn f_{ii}&=&\frac{{2\pi}}{r_i+r_{i-1}+\varrho_i} . \eeqa
Now we
can calculate $\phi=(1_n-\lambda f_x
\lambda^\dagger)^{-1}$. In the following we will be concerned with
the upper-left $2\times 2$ block of $\phi$. Let us denote it by
$\phi_{2 \times 2}$. In this block $1_n$ cancels with the order
$R^{-1}$ terms in $f$.  So \beq \phi_{2\times2} = -(\zeta_i f^{(2)}
\zeta^\dag_j)^{-1} = R \left(\ba{cc} \frac{
         \vec r \cdot \vec \rho_1 +
          r \rho_1 \coth (2 \pi r \nu) }{2} &
          -\frac{\pi r e^{-2 \pi i
         x_0 \nu}\zeta_1 \zeta^\dag_2 }{2       \sinh(2 \pi r        \nu_1) R \rho_{1}       \rho_{2}} \\
     - \frac{\pi r e^{2 \pi i
         x_0 \nu}\zeta_2 \zeta^\dag_1}{2
       \sinh(2 \pi r
        \nu_1) R \rho_{1}
      \rho_{2}} &
        \frac{
         \vec r \cdot \vec \rho_2 +
          r \rho_2 \coth (2 \pi r \nu)}{2}
        \ea \right)^{-1}.
\eeq
It shows that $\phi_{2 \times 2}$ is of order $R$ (other components of $\phi$ are at best of order 1).
As it was shown in \cite{KvBSUN}
\beqa
 {v^1}^m_n &=& -\phi^{-1/2}_{mn}
\label{u_i}, \\
\nn{v^2}^\alpha_m&=&
\left(\frac{\d_z}{2\pi i}-r_\mu\sigma_\mu\right)_\beta^\alpha s^f_k(z)f_{kn}\zeta^{\dag\beta}_n\phi^{1/2}_{nm}\;,
\eeqa
here we denote $\vec r \equiv \vec r_1$ and $\nu\equiv\nu_1$.
In our case only $s_1$ and $s_2$ survive
\beq \label{s12}
s_1(z)=e^{2\pi i x_0(z-\mu_1)}\frac{\sinh[2\pi r(\mu_2-z)]}{\sinh(2\pi r\nu)}\delta_{1[z]}
,\;\;\;\;\;
s_2(z)=e^{2\pi i x_0(z-\mu_2)}\frac{\sinh[2\pi r(z-\mu_1)]}{\sinh(2\pi r\nu)}\delta_{1[z]} .
\eeq
Let us introduce a $2\times 2$ matrix
$u^{\alpha}_\beta\equiv {v^2}^{\alpha}_\beta$.
Using
\beqa \label{Ds1}\label{Ds2}
\left(\frac{\d_z}{2\pi i}-r_\mu\sigma_\mu\right)s_1(z)&=&i r e^{2\pi i x_0\nu}\frac{e^{-2\pi
i r_\mu\sigma^\dag_\mu(\mu_2-z)}}{\sinh(2 \pi r \nu)}\delta_{1[z]}\equiv D\!s_1 ,\\
\left(\frac{\d_z}{2\pi i}-r_\mu\sigma_\mu\right)s_2(z)&=&-i r e^{-2\pi i x_0\nu}\frac{
\nn e^{2\pi i r_\mu\sigma^\dag_\mu(z-\mu_1)}}{\sinh(2 \pi r \nu)}\delta_{1[z]}\equiv D\!s_2
,\eeqa
we get the leading-order expression for $u$ that follows directly from \eq{u_i}:
\beq
u(z)=
e^{2\pi i r_\mu\sigma_\mu^\dag z}\left.\left(
\bea{cc}
{D\!s_1}^1_\gamma{\zeta^{\dag}}^\gamma_1\frac{\pi}{\vrho_1} & {D\!s_2}^1_\gamma{\zeta^{\dag}}^\gamma_2\frac{\pi}{\vrho_2}\\
{D\!s_1}^2_\gamma{\zeta^{\dag}}^\gamma_1\frac{\pi}{\vrho_1} & {D\!s_2}^2_\gamma{\zeta^{\dag}}^\gamma_2\frac{\pi}{\vrho_2}
\eea
\right)\right|_{z=0}\phi^{1/2}_{2\times 2}
.\eeq
Writing the dependence on $x_0$ explicitly we get
\beq
u(z)=
u_{\rm dyon}(\tilde z)U\!\left(\!\!
\bea{cc}
e^{\pi i x_0\nu} & 0\\
0 & e^{-\pi i x_0\nu}
\eea\!\!
\right),\;\;\;\;\;U\!\equiv\! e^{\pi r_i \tau_i(\!\mu_1\!+\!\mu_2\!)}\sqrt{\frac{\sinh(2\pi\nu r)}{2\pi\nu r}}\left[\left(
\bea{cc}
{D\!s_1}^1_\gamma{\zeta^{\dag}}^\gamma_1\frac{\pi}{\vrho_1} & {D\!s_2}^1_\gamma{\zeta^{\dag}}^\gamma_2\frac{\pi}{\vrho_2}\\
{D\!s_1}^2_\gamma{\zeta^{\dag}}^\gamma_1\frac{\pi}{\vrho_1} & {D\!s_2}^2_\gamma{\zeta^{\dag}}^\gamma_2\frac{\pi}{\vrho_2}
\eea
\right)
\!\phi^{1/2}_{2\times 2}\right]_{z,x_0=0}
.\eeq
Here $\tilde z=z-\frac{\mu_1+\mu_2}{2}$ and
\beq
\label{sp1}
 u_{\rm dyon}(r, z)= \sqrt{\frac{2 \pi r}{\sinh(2 \pi \nu
r)}}\exp(2 \pi i r_\mu \sigma^\dag_\mu z).
\eeq
From (\ref{v1v2def}) it follows immediately that $u$ is normalized, i.e.
\beq
\int_{\mu_1}^{\mu_2}u^\dag(z)u(z)\;dz=1_2
,\eeq
it implies that $U$ is a unitary matrix and thus that in the considered limit
the caloron gauge field becomes that of the BPS dyon. The gauge transformation matrix
$U$ gives a connection with the BPS dyon in standard 'hedgehog' gauge. However the
gauge transformation is time dependent. In periodical gauges (see discussion at the end of
Section \ref{pergauges}) we have
\beq
u^{{\rm per}_k}(z)=u_{\rm dyon}(\tilde z) \;U\exp\left[2\pi i x_0\left(\frac{k}{N}+\frac{\mu_1+\mu_2}{2}\right)\right],\;\;\;\;\;k=0,\dots,N-1
\la{uper}
,\eeq
that corresponds to the $SU(2)$ BPS dyon gauge field plus
a constant that comes from the last $U(1)$ factor in \eq{uper}.
I.e. when the separations between dyons are large the gauge field near the $l^{\rm th}$ dyon
 is an almost (with polynomial precision) zero $N\times N$ matrix with only
$2\times 2$ block at $l,l+1$ position filled by the BPS dyon gauge field,
plus a constant diagonal $N\times N$ matrix
\beqa\la{Amunear}
&&A^{ l^{\rm th}{\rm\; block}\;2\times 2}_\mu=A^{\rm dyon}_\mu(\nu_l,\vec x-\vec y_l)+2\pi i\left(\frac{k}{N}+\frac{\mu_l+\mu_{l+1}}{2}\right)1_2\;,\\
\nn &&A^{{\rm outside}\;l^{\rm th}{\rm\; block}\;2\times 2}_\mu=2\pi i\;{\rm diag}\left\{\frac{k}{N}+\mu_1,\dots,\frac{k}{N}+\mu_N\right\}
.\eeqa

It is easy to generalize the result onto the case when only 'neighbor' in
color space dyons are well separated, i.e. when only $\vrho_i\gg 1$.
Then for any $\vec x$ the gauge field will have the same form as in \eq{Amunear}
with $2\times 2$ blocks for each dyon close to the considered point in space $\vec x$.

\end{document}